\documentclass[aip,jcp,reprint]{revtex4-1}
\usepackage{amsfonts}
\usepackage{amsmath}
\usepackage{amssymb}
\usepackage{graphicx}
\usepackage{bm}
\usepackage{epstopdf}

\newtheorem{proposition}{Proposition}

\numberwithin{lemma}{section}
\numberwithin{theorem}{section}
\numberwithin{remark}{section}
\numberwithin{corollary}{section}
\numberwithin{proposition}{section}
\numberwithin{definition}{section}
\numberwithin{example}{section}
\newcommand{\half}{{\textstyle \frac{1}{2}}}

\usepackage{color}
\usepackage{soul}

\newcommand{\Rf}[1]{{#1}}
\newcommand{\Rs}[1]{{#1}}
\newcommand{\our}[1]{{#1}}
\newcommand{\trim}[1]{{#1}}

\begin{document}

\title{New Langevin and Gradient Thermostats for Rigid Body Dynamics}

\author{R.~L. Davidchack}
\email{r.davidchack@le.ac.uk}
\affiliation{Department of Mathematics, University of Leicester, Leicester, LE1~7RH, UK}

\author{T.~E. Ouldridge}
\email{t.ouldridge@imperial.ac.uk}
\affiliation{Department of Mathematics, Imperial College, London SW7~2AZ, UK}

\author{M.~V. Tretyakov}
\email{Michael.Tretyakov@nottingham.ac.uk}
\affiliation{School of Mathematical Sciences, University of Nottingham, Nottingham, NG7~2RD, UK}

\begin{abstract}
We introduce two new thermostats, one of Langevin type and one of gradient (Brownian) type, for rigid body dynamics.
We formulate rotation using the quaternion representation of angular coordinates;
both thermostats preserve the unit length of quaternions.  The Langevin thermostat also
ensures that the conjugate angular momenta stay within the tangent space of the quaternion coordinates,
as required by the Hamiltonian dynamics of rigid bodies.
We have constructed three geometric numerical integrators for the Langevin thermostat and one for the gradient thermostat.
The numerical integrators reflect key properties of the thermostats themselves. Namely, they all preserve the unit length of quaternions, automatically, without the need of a projection onto the unit sphere. The Langevin integrators also ensure that the angular momenta remain within the tangent space of the quaternion coordinates.
The Langevin integrators are quasi-symplectic and of weak order two.  The numerical method for the gradient thermostat is of weak order one.
Its construction exploits ideas of Lie-group type integrators for differential equations on manifolds.
We numerically compare the discretization errors of the Langevin integrators, as well as the efficiency of the gradient integrator compared to the Langevin ones when used in the simulation of rigid TIP4P water model with smoothly truncated electrostatic interactions. We observe that the gradient integrator is computationally less efficient than the Langevin integrators.  We also compare the relative accuracy of the Langevin integrators in evaluating various static quantities and give recommendations as to the choice of an appropriate integrator.


\end{abstract}

\maketitle

\section{Introduction}
In molecular simulations it is often desirable to \trim{fix} the temperature of the simulated system, \trim{ensuring the system samples} from the $NVT$ ensemble (Gibbs measure).
 In molecular dynamics (MD) simulations, the thermostatting is achieved either deterministically (e.g.~Nos\'e-Hoover thermostats) \trim{by} coupling the system to additional degrees of freedom representing a thermal bath \cite{Allen,Ben}, or by a combination of damping and random perturbation of the motion formulated as a stochastic Langevin equation \cite{Schlick,Snook,DHT09,MT7}.  A combination of the deterministic and stochastic approaches is also possible \cite{Bulgac90,Samoletov07,BL10}.

A particular advantage of the Langevin approach is that \trim{all degrees} of freedom of
the system can be thermostated independently, without having to rely on the efficient
energy exchange between \trim{them}.
Good energy exchange is particularly hard to achieve between components
of the system which evolve on different time scales (i.e., fast-slow separation of degrees of freedom).
For Langevin equations, it is easier to ensure and to prove ergodicity (with the Gibbsian invariant measure)
of the thermostat.

Langevin thermostats are also useful for models of dilute \trim{systems} in which the solvent is treated implicitly \cite{Snook}. In the absence of thermostatting, isolated molecules would move ballistically and the energy of the solute would be conserved. Augmenting the Hamiltonian dynamics of the solute with damping and noise leads to diffusive motion and couples the simulated molecules to a thermal reservoir, mimicking some \trim{ solvent effects}.

In systems \trim{that} contain rigid bodies, \trim{approaches must be developed for thermostatting} rotational degrees of freedom.  In our previous work \cite{DHT09}, \trim{we augmented the quaternion-based algorithm of Miller III {\em et al.}\cite{qua02} for  $NVE$ simulation of rigid bodies to include thermostatting}. \Rf{A range of methods for simulation of rigid bodies in the $NVE$ ensemble exist
(see e.g. Refs.~\onlinecite{Evans77a,*Evans77b,DLM97,qua02,Ben,Ome07,*vanZon08jcp,*Omelyan08pre}
and references therein).
We choose to build upon the work of Miller III {\em et al.}~because their numerical integrator is widely used, symplectic, preserves the length of quaternions exactly and is simple to implement}. In Ref.~\onlinecite{DHT09} using the operator splitting approach, two different weak 2nd-order (i.e., with 2nd-order convergence of approximate averages to the exact ones; see further details on stochastic numerics, for example, in Refs.~\onlinecite{MT1,MT7}) methods were presented and tested.  They were combined with either Langevin or gradient dynamics for the translational degrees of freedom.
\trim{Subsequently}, this method was applied to simulate a coarse-grained, implicit-solvent model of DNA in a range of contexts.\cite{Ouldridge_walker_2013,Ouldridge_binding_2013,Srinivas2013,Machinek2014}

\Rf{
Alternative methods for Langevin thermostats that can be used for rigid bodies were considered in Refs.~\onlinecite{Eric06,Sun08}. In Ref.~\onlinecite{Eric06} second-order integrators for Langevin equations with holonomic constraints were proposed. These integrators use SHAKE to deal with holonomic constraints which is computationally expensive in the specific case of rigid bodies (it is so even in the case of deterministic sampling from $NVE$ ensembles\cite{qua02,Ome07}). In Ref.~\onlinecite{Sun08} the authors consider Langevin dynamics for rigid bodies using the rotational integration scheme based on the Lie--Poisson integrator\cite{DLM97,Ben}. It is slightly more expensive than the quaternion representation which is the minimal non-singular description of rotations.
}

In this work we first \trim{propose a new  Langevin thermostat for rigid body dynamics.  This thermostat improves upon that in Ref.~\onlinecite{DHT09},
as it not only preserves the unit length of quaternions but {\it also} keeps the
angular momenta conjugate to the quaternion coordinates on the tangent space.
Although the latter does not hold for the old thermostat of  Ref.~\onlinecite{DHT09}, we show that this non-physical behaviour
does not affect quantities of physical interest; it can, however, introduce rounding errors
in numerical integration.} For the new thermostat, we construct three geometric integrators (Langevin A, B and C) of weak order two; \trim{Langevin A and B} are related to \trim{those} in Ref.~\onlinecite{DHT09}, whereas Langevin C is qualitatively distinct. Langevin C has accuracy similar to
Langevin A but, like Langevin B, can be used with
large values of the friction parameters (i.e.~approaching the over-damped limit).
\trim{We also simplify the construction of integrators,
eliminating the explicit matrix exponentials and Cholesky decompositions  required in Ref.~\onlinecite{DHT09}}.
Further, we propose a new gradient (Brownian) thermostat for rigid body dynamics and
construct a 1st-order geometric integrator for it.  Both the gradient thermostat and the numerical
scheme preserve the unit length of quaternions.  We perform numerical comparison
of the proposed Langevin and gradient thermostats and of the derived numerical integrators.
\trim{These tests demonstrate that the Langevin thermostat integrated by Langevin A or Langevin C is a powerful approach  for computing $NVT$
ensemble averages involving rigid bodies}.

The rest of the paper is organised as follows. \trim{Having defined  important quantities in Section~\ref{sec:old}, we present a new Langevin thermostat for rigid body dynamics in Section~\ref{sec:new}, contrasting it with that of   Ref.~\onlinecite{DHT09}. A new gradient (Brownian) thermostat is introduced in Section~\ref{sec:grad}.
In Section~\ref{sec:met} we derive numerical integrators for the thermostats of Sections~\ref{sec:new} and \ref{sec:grad}.
In Section~\ref{sec:num} we perform numerical tests of the proposed integrators using a screened TIP4P rigid water model.
Comparative performance of the thermostats and numerical methods for them is discussed in Section~\ref{sec:concl}.}

\section{Preliminaries\label{sec:old}}
We \trim{consider $n$ rigid three-dimensional bodies with} center-of-mass coordinates $\mathbf{r}=(r^{1\mathsf{T}},\ldots ,r^{n\,%
\mathsf{T}})^{\mathsf{T}}\in \mathbb{R}^{3n},$ $%
r^{j}=(r_{1}^{j},r_{2}^{j},r_{3}^{j})^{\mathsf{T}}\in \mathbb{R}^{3},$ and
the rotational coordinates in the quaternion representation $\mathbf{q}%
=(q^{1\,\mathsf{T}},\ldots ,q^{n\,\mathsf{T}})^{\mathsf{T}}$, $%
q^{j}=(q_{0}^{j},q_{1}^{j},q_{2}^{j},q_{3}^{j})^{\mathsf{T}},$ such that
$|q^{j}|=1$, i.e., $q^{j}\in \mathbb{S}^{3}$, which is the
three-dimensional unit sphere with center at the origin.
We use standard matrix notations, and ``$\mathsf{T}$'' denotes transpose.
For further background on the quaternion representation of rigid body dynamics, \trim{see Refs.~\onlinecite{Evans77a,*Evans77b,quabook,quabook2}}.

Following Ref.~\onlinecite{qua02}, we write the system Hamiltonian \trim{as}
\begin{equation}
H(\mathbf{r},\mathbf{p},\mathbf{q},\bm{\pi})=\frac{\mathbf{p}^{\mathsf{T}}%
\mathbf{p}}{2m}+\sum_{j=1}^{n}\sum_{l=1}^{3}V_{l}(q^{j},\pi ^{j})+U(\mathbf{r%
},\mathbf{q}),  \label{a1}
\end{equation}%
where $\mathbf{p}=(p^{1\,\mathsf{T}},\ldots ,p^{n\,\mathsf{T}})^{\mathsf{T}%
}\in \mathbb{R}^{3n}$, $p^{j}=(p_{1}^{j},p_{2}^{j},p_{3}^{j})^{\mathsf{T}%
}\in \mathbb{R}^{3},$ are the center-of-mass momenta conjugate to $\mathbf{r}
$; $\bm{\pi}=(\pi ^{1\,\mathsf{T}},\ldots ,\pi ^{n\,\mathsf{T}})^{\mathsf{T}%
} $, $\pi ^{j}=(\pi _{0}^{j},\pi _{1}^{j},\pi _{2}^{j},\pi _{3}^{j})^{%
\mathsf{T}}$ are the angular momenta conjugate to $\mathbf{q}$ {such that $%
q^{j\,\mathsf{T}}\pi ^{j}=0,$ i.e., $\pi ^{j}\in T_{q^{j}}\mathbb{S}^{3}$,
which is the tangent space of $\mathbb{S}^{3}$ at $q^{j};$} and $U(\mathbf{r}%
,\mathbf{q})$ is the \trim{potential energy}. The second term in (\ref%
{a1}) represents the rotational kinetic energy, \trim{with}
\begin{equation}
V_{l}(q,\pi )=\frac{1}{8I_{l}}\left[ \pi ^{\mathsf{T}}S_{l}q\right] ^{2},\ \
l=1,2,3,  \label{a2}
\end{equation}%
where the three constant $4$-by-$4$ matrices $S_{l}$ are
\begin{eqnarray*}
S_{1} &=&\left[
\begin{array}{cccc}
0 & -1 & 0 & 0 \\
1 & 0 & 0 & 0 \\
0 & 0 & 0 & 1 \\
0 & 0 & -1 & 0%
\end{array}%
\right] ,\ S_{2}=\left[
\begin{array}{cccc}
0 & 0 & -1 & 0 \\
0 & 0 & 0 & -1 \\
1 & 0 & 0 & 0 \\
0 & 1 & 0 & 0%
\end{array}%
\right], \\
S_{3} &=&\left[
\begin{array}{cccc}
0 & 0 & 0 & -1 \\
0 & 0 & 1 & 0 \\
0 & -1 & 0 & 0 \\
1 & 0 & 0 & 0%
\end{array}%
\right],
\end{eqnarray*}%
and $I_{l}$ are the principal moments of inertia of the \trim{molecule}. We also
introduce $S_{0}=\mbox{diag}(1,1,1,1),$ the \trim{matrix}
$D=\mbox{diag}(0,1/I_{1},1/I_{2},1/I_{3})$, and the orthogonal matrix:
\begin{eqnarray*}
S(q)=[S_{0}q,S_{1}q,S_{2}q,S_{3}q]
=\left[
\begin{array}{cccc}
q_{0} & -q_{1} & -q_{2} & -q_{3} \\
q_{1} & q_{0} & -q_{3} & q_{2} \\
q_{2} & q_{3} & q_{0} & -q_{1} \\
q_{3} & -q_{2} & q_{1} & q_{0}%
\end{array}%
\right] .
\end{eqnarray*}%
Note that $q^{\mathsf{T}}S(q)=(1,0,0,0)$ and $q^{\mathsf{T}}S(q)D=(0,0,0,0)$.
The rotational kinetic energy of a molecule can be expressed in terms of the matrices $D$ and $S$ as follows:
\[ \sum_{l=1}^{3}V_{l}(q,\pi) = \frac{1}{8}\pi^\mathsf{T} S(q) D S^\mathsf{T}(q) \pi\,. \]

We assume that $U(\mathbf{r},\mathbf{q})$ is a sufficiently smooth function.
Let $f${$^{j}(\mathbf{r},\mathbf{q})=-\nabla_{r^{j}}U(\mathbf{r},\mathbf{q})\in \mathbb{R}^{3}$, \trim{the net force acting on molecule} $j$, and $F^{j}(%
\mathbf{r},\mathbf{q})=-\tilde{\nabla}_{q^{j}}$}$U(\mathbf{r},\mathbf{q})\in
T_{q^{j}}\mathbb{S}^{3},$ which is the rotational force.  Note that, while
$\nabla _{r^{j}}$ is the gradient in the Cartesian coordinates in
$\mathbb{R}^{3}$, $\tilde{\nabla}_{q^{j}}$ is the directional derivative
\footnote{Note that in Eq.~(3) of Ref.~\onlinecite{DHT09} the notation $\nabla
_{q^{j}}$ for the directional derivative was used in $\nabla _{q^{j}}U$
while $\nabla _{q^{j}}V$ meant the conventional gradient.} tangent to the
three dimensional sphere $\mathbb{S}^{3}$ implying that
\begin{equation}
\mathbf{q}^{\mathsf{T}}\tilde{\nabla}_{q^{j}}U(\mathbf{r},\mathbf{q})=0.
\label{qU0}
\end{equation}%
\trim{The generalized force $F^{j}(\mathbf{r},\mathbf{q})$ can be calculated directly or via
Cartesian torques: see Appendix~\ref{sec:forces}. The derivatives of (\ref{a1}) determine the
dynamics without damping and noise. In particular, we note}
\begin{eqnarray} \label{matrixrel}
\sum_{l=1}^{3}\nabla _{\pi }V_{l}(q,\pi ) &=&\frac{1}{4}\sum_{l=1}^{3}\frac{1%
}{I_{l}}S_{l}q\left[ S_{l}q\right] ^{\mathsf{T}}\pi \\
&=&\frac{1}{4} S(q) DS^\mathsf{T}(q)\pi ,  \notag \\
\sum_{l=1}^{3}\nabla _{q}V_{l}(q,\pi ) &=&-\frac{1}{4}\sum_{l=1}^{3}\frac{1}{%
I_{l}}\left[ \pi ^{\mathsf{T}}S_{l}q\right] S_{l}\pi . \notag
\end{eqnarray}

\section{New Langevin thermostat for rigid body dynamics\label{sec:new}}
\trim{We propose the following Langevin thermostat for rigid body dynamics:
\begin{eqnarray}
dR^{j} &=&\frac{P^{j}}{m}dt,\ \ R^{j}(0)=r^{j},  \label{nl1} \\
dP^{j} &=&f{^{j}(\mathbf{R},\mathbf{Q})}dt  \notag \\
&&-\gamma P^{j}dt+\sqrt{\frac{2m\gamma }{\beta }}dw^{j}(t),\ \
P^{j}(0)=p^{j},  \notag \\
dQ^{j} &=&\frac{1}{4}S(Q^{j})DS^\mathsf{T}(Q^{j})\Pi ^{j}dt,\
Q^{j}(0)=q^{j},  \label{nl2} \\  &&  |q^{j}|=1, \notag \\
d\Pi ^{j} &=&\frac{1}{4}\sum_{l=1}^{3}\frac{1}{I_{l}}
\left(\Pi^{j\,\mathsf{T}}S_{l}Q^{j}\right) S_{l}\Pi^{j}dt+F{^{j}(\mathbf{R},\mathbf{Q})}dt
\notag \\
&&-\Gamma J(Q^{j})\Pi ^{j}dt+\sqrt{\frac{2M\Gamma }{\beta }}%
\sum_{l=1}^{3}S_{l}Q^{j}dW_{l}^{j}(t),\notag \\
&&\Pi ^{j}(0)=\pi ^{j},\ \ {q^{j\,\mathsf{T}}\pi ^{j}=0},\ \ j = 1,\ldots ,n,  \notag
\end{eqnarray}%
where $(\mathbf{w}^{\mathsf{T}},\mathbf{W}^{\mathsf{T}})^{\mathsf{T}}=(w^{1\,\mathsf{T}},\ldots ,w^{n\,\mathsf{T}},W^{1\,\mathsf{T}},\ldots ,$ $W^{n\,%
\mathsf{T}})^{\mathsf{T}}$ is a $(3n+3n)$-dimensional standard Wiener
process with $w^{j}=(w_{1}^{j},w_{2}^{j},w_{3}^{j})^{\mathsf{T}}$ and $%
W^{j}=(W_{1}^{j},W_{2}^{j},W_{3}^{j})^{\mathsf{T}}$;  $\gamma \geq 0$
and $\Gamma \geq 0$ are the friction coefficients for the translational
and rotational motions and  $\beta =1/(k_{B}T)>0$ is the inverse temperature.} In the above equations we also use
\begin{equation}
J(q)={\frac{M}{4}S(q)DS^\mathsf{T}(q),}\ \text{\ }M=\frac{4}{\sum_{l=1}^{3}%
\frac{1}{I_{l}}}.  \label{a15}
\end{equation}

It is not difficult to show that the new thermostat (\ref{nl1})--(\ref{nl2}) possesses the following
properties:

\begin{itemize}
\item The Ito interpretation of the system of stochastic differential equations (SDEs) (\ref{nl1})--(\ref{nl2})
coincides with its Stratonovich interpretation.

\item \trim{The} solution of (\ref{nl1})--(\ref{nl2}) preserves the quaternion length 
\begin{equation}
|Q^{j}(t)|=1,\ \ j=1,\ldots ,n\,,\ \ \mbox{for all~}t\geq 0.  \label{a211}
\end{equation}%

\item The solution of (\ref{nl1})--(\ref{nl2}) automatically preserves the
following constraint:
\begin{equation}
\mathbf{Q}^{\mathsf{T}}(t)\mathbf{\Pi }(t)=0,\ \ \mbox{for all~}t\geq 0.
\label{qpi0}
\end{equation}
\trim{Physically, $Q^{j}(t)$ are constrained to unit spheres; therefore the momenta $\Pi^{j}(t)$ should also have three degrees of freedom. This constraint is manifest in the physical requirement $\mathbf{Q}^{\mathsf{T}}(t)\mathbf{\Pi }(t)=0$, i.e., $\Pi^{j}(t)\in T_{Q^{j}}\mathbb{S}^{3}$.}

\item Assume that the solution $X(t)=(\mathbf{R}^{\mathsf{T}}(t),$ $\mathbf{P}^{%
\mathsf{T}}(t),\mathbf{Q}^{\mathsf{T}}(t),\mathbf{\Pi }^{\mathsf{T}}(t))^{%
\mathsf{T}}$ of (\ref{nl1})--(\ref{nl2}) is an ergodic process \cite{Has,Soize} on
\begin{eqnarray*}
\ \ \mathbb{D}&=&\{x=(\mathbf{r}^{\mathsf{T}},\mathbf{p}^{\mathsf{T}},\mathbf{q}^{%
\mathsf{T}},\bm{\pi}^{\mathsf{T}})^{\mathsf{T}}\in \mathbb{R}^{14n}:\\
&& \ \ |q^{j}|=1,\ \ q^{j\,\mathsf{T}}\pi^{j}=0,\ \ j=1,\ldots ,n\}.
\end{eqnarray*}%
Then it can be shown that the invariant measure of $X(t)$ is Gibbsian with the density $\rho (\mathbf{r},
\mathbf{p},\mathbf{q},\bm{\pi})$ on $\mathbb{D}$:
\begin{equation}\label{a3}
\rho (\mathbf{r},\mathbf{p},\mathbf{q},\bm{\pi})\varpropto \exp (-\beta H(%
\mathbf{r},\mathbf{p},\mathbf{q},\bm{\pi })),
\end{equation}%
which corresponds to the $NVT$ ensemble of rigid bodies, as required.
\end{itemize}

\trim{The Langevin thermostat of Ref.~\onlinecite{DHT09} is similar to (\ref{nl1})--(\ref{nl2}) (see Appendix~\ref{app:old}). However, although it preserves (\ref{a211}) the condition (\ref{qpi0}) is violated. As we show in Appendix~\ref{app:old}, the unphysical component of angular momentum for each particle,  $\Omega ^{j}(t):=\half(Q^{j}(t))^{\mathsf{T}}\Pi ^{j}(t)$, has a variance that grows linearly in time for the thermostat of Ref.~\onlinecite{DHT09}: $E\left[\Omega^{j}(t)\right]^{2}=M\Gamma t/(2\beta)$. In fact, $\Omega^{j}(t) \neq 0$ does not affect evaluation of quantities of physical interest (see Appendix~\ref{sec:just}), and hence they are computed correctly using the thermostat of Ref.~\onlinecite{DHT09}. At the same time, unbounded growth of the variance of this non-physical component  can introduce rounding errors during numerical integration. The new thermostat does not have this deficiency. We also note that the old thermostat of Ref.~\onlinecite{DHT09} required a $7n$-dimensional Wiener process while the new thermostat
(\ref{nl1})--(\ref{nl2}) requires a $6n$-dimensional Wiener process, which is consistent with the number of degrees of freedom in the system.}

\section{Gradient thermostat for rigid body dynamics\label{sec:grad}}
Gradient systems are popular in molecular dynamics for thermostatting
translational degrees of freedom \cite{Doll87,Schlick,MT7} (see also
references therein). In Ref.~\onlinecite{DHT09} a mixture of a gradient system for the
translational dynamics and a Langevin-type equation for the rotational
dynamics was suggested. In this Section we propose Brownian dynamics for
thermostatting rigid \trim{bodies, i.e., a gradient system for
center-of-mass and rotational coordinates}.

It is easy to verify that
\begin{eqnarray}
\int_{\mathbb{D}_\mathrm{mom}} \exp && (-\beta H(\mathbf{r},\mathbf{p},\mathbf{q},
\bm{\pi }))d\mathbf{p}d\bm{\pi } \label{newden} \\ && \varpropto \exp (-\beta U(\mathbf{r},\mathbf{q}))
=:\tilde{\rho}(\mathbf{r},\mathbf{q}), \notag
\end{eqnarray}%
where $(\mathbf{r}^{\mathsf{T}},\mathbf{q}^{\mathsf{T}})^{\mathsf{T}}
\in \mathbb{D}^{\prime}=\{(\mathbf{r}^{\mathsf{T}},\mathbf{q}^{\mathsf{T}%
})^{\mathsf{T}}\in \mathbb{R}^{7n}:\ \ |q^{j}|=1\}$ and the domain of conjugate
momenta $\mathbb{D}_\mathrm{mom}=\{(\mathbf{p^{\mathsf{T}},}\bm{\pi }^{\mathsf{T}})^{\mathsf{T}}\in
\mathbb{R}^{7n}:\mathbf{q}^{\mathsf{T}}\bm{\pi}=0\}.$

We introduce the gradient system in the form of Stratonovich SDEs:
\begin{align}
d\mathbf{R}& =\frac{\upsilon }{m}\mathbf{f}(\mathbf{R},\mathbf{Q})dt+\sqrt{\frac{%
2\upsilon }{m\beta }}d\mathbf{w}(t),\ \ \mathbf{R}(0)=\mathbf{r},  \label{a10} \\
dQ^{j}& =\frac{\Upsilon }{M}F{^{j}(\mathbf{R},\mathbf{Q})}dt+\sqrt{\frac{%
2\Upsilon }{M\beta }}\sum_{l=1}^{3}S_{l}Q^{j}\star dW_{l}^{j}(t),\ \
\label{a100} \\
&Q^{j}(0)=q^{j},\ \ |q^{j}|=1,\ \ j=1,\ldots ,n,  \notag
\end{align}%
where ``$\star$'' indicates the Stratonovich form of the SDEs,
parameters $\upsilon >0$ and $\Upsilon >0$ control the speed of
evolution of the gradient system (\ref{a10})--(\ref{a100}),
$\mathbf{f}=(f^{1\,\mathsf{T}},\ldots ,f^{n\,\mathsf{T}})^{\mathsf{T}}$
and the rest of the notation is as in (\ref{nl1})--(\ref{nl2}).
Note that, unlike in the case of (\ref{nl1})--(\ref{nl2}), the
Stratonovich and Ito interpretations of the SDEs (\ref{a10})--(\ref{a100}) do
not coincide.

This new gradient thermostat possesses the following properties.

\begin{itemize}
\item As in the case of (\ref{nl1})--(\ref{nl2}),
the solution of (\ref{a10})--(\ref{a100}) preserves the quaternion length (\ref{a211}).

\item Assume that the solution $X(t)=(\mathbf{R}^{\mathsf{T}}(t),\mathbf{Q}^{%
\mathsf{T}}(t))^{\mathsf{T}}\in \mathbb{D}^{\prime }$ of (\ref{a10})--(\ref%
{a100}) is an ergodic process \cite{Has}. Then, by the usual means of the stationary
Fokker-Planck equation (see Appendix~\ref{sec:fpe}), one can show that its
invariant measure is Gibbsian with the density $\tilde{\rho}(\mathbf{r},%
\mathbf{q})$ from (\ref{newden}).
\end{itemize}

\trim{If a thermostat is used only to control the temperature of a system,
and not to mimic the dynamical effects of an implicit solvent, 
the Langevin thermostat (\ref{nl1})--(\ref{nl2}) is
suitable for computing both dynamical and static quantities
(provided the friction coefficients are relatively small). By contrast, the
gradient thermostat (\ref{a10})--(\ref{a100}) can be used to compute only
static quantities \cite{Schlick,MT7} in such systems.}

\section{Numerical methods \label{sec:met}}
In this section we construct geometric integrators for the new Langevin
thermostat (\ref{nl1})--(\ref{nl2}) (Sections~\ref{sec:LA}-\ref{sec:LC}) and
for the new gradient thermostat (\ref{a10})--(\ref{a100}) (Section~\ref{sec:Ngt}).
The numerical methods for the Langevin thermostat are based on the splitting technique.
It was observed in Ref.~\onlinecite{BenM13} that numerical
schemes based on different splittings might have considerably different
properties.  Roughly speaking, Langevin thermostat SDEs (\ref{nl1})--(\ref{nl2})
consist of three components: Hamiltonian $+$ damping $+$ noise.
The integrator Langevin A is based on the splitting of (\ref{nl1})--(\ref{nl2}) into
\our{a system close to a stochastic Hamiltonian system} (Hamiltonian $+$ noise) and the
deterministic system of linear differential equations corresponding to the
Langevin damping.  The other two schemes, Langevin~B and C, are based on
splitting of (\ref{nl1})--(\ref{nl2}) into the deterministic Hamiltonian system
and the Ornstein-Uhlenbeck process (damping $+$ noise) using their different concatenations.
All three schemes are of weak order 2 and use one evaluation of forces per step.
The numerical method for the gradient thermostat (\ref{a10})--(\ref{a100}) also
uses one force evaluation per step, but it is of weak order 1. To preserve the length of
quaternions in the case of numerical integration of (\ref{a10})--(\ref{a100}),
we use ideas of Lie-group type integrators for differential equations on manifolds
(see, for example, Ref.~\onlinecite{HLW02}).

In what follows we assume that (\ref{nl1})--(\ref{nl2}) and (\ref{a10})--(\ref{a100})
have to be solved on a time interval $[0,T]$ and, for simplicity, we
use a uniform time discretization with the step $h=T/N$.

In Secs.~\ref{sec:LA}-\ref{sec:LC} we use the mapping $\Psi _{t,l}(q,\pi
):$ $(q,\pi )\mapsto (\mathcal{Q},\mathit{\Pi })$ defined by (see, e.g.
Refs.~\onlinecite{qua02,DHT09}):
\begin{equation}
\begin{split}
\mathcal{Q}& =\cos (\chi _{l}t)q+\sin (\chi _{l}t)S_{l}q\,, \\
\mathit{\Pi }& =\cos (\chi _{l}t)\pi +\sin (\chi _{l}t)S_{l}\pi \,,
\end{split}
\label{a24}
\end{equation}%
where
\begin{equation*}
\chi _{l}=\frac{1}{4I_{l}}\pi ^{\mathsf{T}}S_{l}q\,.
\end{equation*}%
We also introduce a composite map
\begin{equation}
\Psi_{t}=\Psi_{t/2,3}\circ \Psi_{t/2,2}\circ \Psi_{t,1}\circ \Psi_{t/2,2}
\circ \Psi_{t/2,3}\,, \label{a25}
\end{equation}
where ``$\circ$'' denotes function composition, i.e., $(g\circ f)(x) = g(f(x))$.

\Rf{We note that Langevin A, B and C degenerate to the deterministic symplectic integrator from Ref.~\onlinecite{qua02}
when the Langevin thermostat (\ref{nl1})--(\ref{nl2}) degenerates to the deterministic Hamiltonian system (i.e, when the thermostat is ``switched off").
We base the Langevin methods on the scheme from Ref.~\onlinecite{qua02} because it is symplectic and preserves the length
of quaternions automatically. It is also widely used by the molecular dynamics community and, in particular, implemented
in molecular dynamics libraries, which makes implementation of Langevin A, B and C simpler and more practical.
}

\subsection{Geometric integrator Langevin A: revisited\label{sec:LA}}
The geometric integrator of this section for solving the new Langevin
thermostat (\ref{nl1})--(\ref{nl2}) is similar to the Langevin~A method 
proposed in Ref.~\onlinecite{DHT09} for (\ref{nl1}),~(\ref{lt2}). 
It is based on splitting the Langevin system (\ref{nl1})--(\ref{nl2}) into the
\our{system (\ref{nl1})--(\ref{nl2}) without the damping term,
which is close to a stochastic Hamiltonian system,} and the
deterministic system of linear differential equations
\begin{equation}
\begin{split}
\dot{\mathbf{p}}& =-\gamma \mathbf{p} \\
\dot{\pi}^{j}& =-\Gamma J(q^{j})\pi ^{j},\ j=1,\ldots ,n\,.
\end{split}
\label{nsl1}
\end{equation}%
We construct a weak 2nd-order quasi-symplectic integrator
\our{(i.e., the scheme becomes symplectic when $\gamma = \Gamma=0$) for
the system (\ref{nl1})--(\ref{nl2}) without the damping term}\cite{MilRT02,MT1} and appropriately
concatenate \cite{MT03,MT1} it with the exact solution of (\ref{nsl1}). The
resulting numerical method has the form:%
\begin{eqnarray}
\mathbf{P}_{0} &=&\mathbf{p},\ \ \mathbf{R}_{0}=\mathbf{r},\   \label{firla}
\\
\mathbf{Q}_{0} &=&\mathbf{q}\text{ with }|q^{j}|=1,\ j=1,\ldots ,n,\   \notag
\\
\mathbf{\Pi }_{0} &=&\mathbf{\pi }\text{ with }\mathbf{q}^{\mathsf{T}}\mathbf{\pi
=0,}  \notag \\
\mathcal{P}_{1,k} &=&{\rm e}^{-\gamma\frac{h}{2}}\mathbf{P}_{k}\,,  \notag \\
\mathit{\Pi }_{1,k}^{j} &=& {\rm e}^{-\Gamma J(Q_{k}^{j})\frac{h}{2}}\Pi
_{k}^{j},\ \ j=1,\ldots ,n,  \notag
\end{eqnarray}

\begin{eqnarray*}
\mathcal{P}_{2,k} &=&\mathcal{P}_{1,k}+\frac{h}{2}\mathbf{f}(\mathbf{R}_{k},%
\mathbf{Q}_{k})+\frac{\sqrt{h}}{2}\sqrt{\frac{2m\gamma }{\beta }}\mathbf{\xi
}_{k} \\
\mathit{\Pi }_{2,k}^{j} &=&\mathit{\Pi }_{1,k}^{j}+\frac{h}{2}F{^{j}}(%
\mathbf{R}_{k},\mathbf{Q}_{k})  \\
&&+\frac{\sqrt{h}}{2}\sqrt{\frac{2M\Gamma }{%
\beta }}\sum_{l=1}^{3}S_{l}\mathbf{Q}_{k}\eta _{k}^{j,l}, \\
&&j=1,\ldots ,n, \\
\mathbf{R}_{k+1} &=&\mathbf{R}_{k}+\frac{h}{m}\mathcal{P}_{2,k},
\end{eqnarray*}%
\begin{equation*}
(Q_{k+1}^{j},\mathit{\Pi }_{3,k}^{j})=\Psi _{h}(Q_{k}^{j},\mathit{\Pi }%
_{2,k}^{j}),\ \ j=1,\ldots ,n,
\end{equation*}%
\begin{eqnarray}
\mathit{\Pi }_{4,k}^{j} &=&\mathit{\Pi }_{3,k}^{j}+\frac{h}{2}F{^{j}}(%
\mathbf{R}_{k+1},\mathbf{Q}_{k+1})  \notag \\
&&+\frac{\sqrt{h}}{2}\sqrt{\frac{2M\Gamma }{\beta }}\sum_{l=1}^{3}S_{l}%
\mathbf{Q}_{k+1}\eta _{k}^{j,l},\ j=1,\ldots ,n,  \notag \\
\mathcal{P}_{3,k} &=&\mathcal{P}_{2,k}+\frac{h}{2}\mathbf{f}(\mathbf{R}%
_{k+1},\mathbf{Q}_{k+1})+\frac{\sqrt{h}}{2}\sqrt{\frac{2m\gamma }{\beta }}%
\mathbf{\xi }_{k},  \notag
\end{eqnarray}%
\begin{eqnarray*}
\mathbf{P}_{k+1} &=&{\rm e}^{-\gamma \frac{h}{2}}\mathcal{P}_{3,k}\,, \\
\Pi _{k+1}^{j} &=&{\rm e}^{-\Gamma J(Q_{k+1}^{j})\frac{h}{2}}
\mathit{\Pi}_{4,k}^{j},\ \ j=1,\ldots ,n, \\
k &=&0,\ldots ,N-1,
\end{eqnarray*}%
where $\mathbf{\xi }_{k}=(\xi _{1,k},\ldots ,\xi _{3n,k})^{\mathsf{T}}$ and $%
\eta _{k}^{j,l},$ $l=1,2,3,$ $j=1,\ldots ,n,$ with their components being
i.i.d.~(independent and identically distributed) with the same probability distribution
\begin{equation}
P(\theta =0)=2/3,\ \ P(\theta =\pm \sqrt{3})=1/6.  \label{n31}
\end{equation}

We proved (the proof is not presented here) that the geometric integrator (\ref{firla})--(\ref{n31}) possesses
properties stated in the next proposition. The concept of quasi-symplectic
methods is described in Refs.~\onlinecite{MT03,MT1} and
also Refs.~\onlinecite{MT7,DHT09}. The proof of weak convergence order is done by standard arguments based on
the general convergence theorem (see p.~100 in Ref.~\onlinecite{MT1}).

\begin{proposition}
\label{prp1}The numerical scheme $(\ref{firla})$--$(\ref{n31})$ for 
$(\ref{nl1})$--$(\ref{nl2})$ is quasi-symplectic, it preserves the structural
properties $(\ref{a211})$ and $(\ref{qpi0})$ and it is of weak order two.
\end{proposition}

We note that one can choose $\mathbf{\xi }_{k}$ and $\eta _{k}^{j,l},$ $%
l=1,2,3,$ $j=1,\ldots ,n,$ so that their components are i.i.d.~Gaussian
random variables with zero mean and unit variance. In this case the weak
order of the scheme remains second as \trim{for} the simple discrete
distribution (\ref{n31}). Let us remark in passing that in the case of
Gaussian random variables the above scheme also converges in the mean-square
(also called strong) sense \cite{MT1} with order one.

\trim{Writing $J(q)$} explicitly in terms of $S(q)$ and $D$, see (\ref{a15}),
reveals that the exponent appearing in (\ref{firla}) is easier to compute
than originally suggested in Ref.~\onlinecite{DHT09}. In particular,
explicit evaluation of a matrix exponent is not required:
\begin{eqnarray}
{\rm e}^{-\Gamma J(q)\frac{h}{2}}
&=&{S(q) {\rm e}^{-\frac{\Gamma Mh}{8} D} S^\mathsf{T}(q)} \notag\\
&=&{\sum_{l=1}^{3}{\rm e}^{-\frac{\Gamma Mh}{8I_{l}}}S_{l}q\left[
S_{l}q\right]^{\mathsf{T}}}. \label{expcal}
\end{eqnarray}

\Rs{We note that integrators for the Langevin thermostat
(\ref{nl1})--(\ref{nl2}) can be constructed with the same splitting into damping and Hamiltonian plus noise, but which
use distinct concatenations of these flows.
In the case of Langevin equations for translational degrees of freedom other concatenations were considered, e.g. in
Ref.~\onlinecite{MT03,MT1,Referee2}. Thorough comparison of these methods, even without rotational motion, deserves a separate study.
}

\subsection{Geometric integrator Langevin B: revisited\label{sec:LB}}

The geometric integrator of this section for solving the new Langevin
thermostat (\ref{nl1})--(\ref{nl2}) is similar to the proposed in Ref.~\onlinecite{DHT09}
Langevin~B for (\ref{nl1}),~(\ref{lt2}). It is based on the following
splitting:
\begin{equation}
\begin{split}
d\mathbf{P}_{I}& =-\gamma \mathbf{P}_{I}\,dt+\sqrt{\frac{2m\gamma }{\beta }}d%
\mathbf{w}(t), \\
d\Pi _{I}^{j}& =-\Gamma J(q)\Pi _{I}^{j}dt+\sqrt{\frac{2M\Gamma }{\beta }}%
\sum_{l=1}^{3}S_{l}qdW_{l}^{j}(t);
\end{split}
\label{lb1}
\end{equation}
\begin{equation}
\begin{split}
d\mathbf{R}_{II}=& \frac{\mathbf{P}_{II}}{m}\,dt \\
d\mathbf{P}_{II}=& \mathbf{f}(\mathbf{R}_{II},\mathbf{Q}_{II})dt, \\
dQ_{II}^{j}=& \frac{1}{4}S(Q_{II}^{j})DS^\mathsf{T}(Q_{II}^{j})\Pi
_{II}^{j}dt\,, \\
d\Pi _{II}^{j}=& F{^{j}}(\mathbf{R}_{II},\mathbf{Q}_{II})dt \\
& +\frac{1}{4}\sum_{l=1}^{3}\frac{1}{I_{l}}\left[ (\Pi _{II}^{j})^{\mathsf{T}%
}S_{l}Q_{II}^{j}\right] S_{l}\Pi _{II}^{j}dt\,, \\
j=& 1,\ldots ,n.
\end{split}
\label{lb2}
\end{equation}%
The SDEs (\ref{lb1}) have the exact solution,
\begin{equation}
\begin{split}
\mathbf{P}_{I}(t)& =\mathbf{P}_{I}(0){\rm e}^{-\gamma t}+\sqrt{\frac{2m\gamma }{%
\beta }}\int_{0}^{t}{\rm e}^{-\gamma (t-s)}d\mathbf{w}(s), \\
\Pi _{I}^{j}(t)& ={\rm e}^{-\Gamma J(q)t}\Pi _{I}^{j}(0) \\
& +\sqrt{\frac{2M\Gamma }{\beta }}\sum\limits_{l=1}^{3}\int_{0}^{t}
{\rm e}^{-\Gamma J(q)(t-s)} S_{l}qdW_{l}^{j}(s).
\end{split}
\label{lbe}
\end{equation}
\Rf{Such splittings in the case of Langevin equations for translational degrees of freedom
were considered in e.g. Ref.~\onlinecite{Skeel99,MT03,MT1,BenM13} (see also references therein).}

To construct a method based on the splitting (\ref{lb1})--(\ref{lb2}), we
take half a step of (\ref{lb1}) using (\ref{lbe}), one step of the
symplectic method for (\ref{lb2}) from Ref.~\onlinecite{qua02}, and again half a step
of (\ref{lb1}). \our{Note that Langevin~C in Sec.~\ref{sec:LC} uses 
the same splitting (\ref{lb1})--(\ref{lb2}) but a different concatenation.}

The vector $\int_{0}^{t}{\rm e}^{-\Gamma J(q)(t-s)} S_{l}qdW_{l}^{j}(s)$
in (\ref{lbe}) is Gaussian with zero mean and covariance
\begin{equation*}
C_{l}(t;q)=\int_{0}^{t} {\rm e}^{-\Gamma J(q)(t-s)}S_{l}q(S_{l}q)^{\mathsf{T}} {\rm e}^{-\Gamma J(q)(t-s)}ds.
\end{equation*}
\trim{The} covariance matrix $C(t;q)$ of the sum $%
\sum_{l=1}^{3}\int_{0}^{t}{\rm e}^{-\Gamma J(q)(t-s)}S_{l}qdW_{l}^{j}(s)$ is
equal to
\begin{equation*}
C(t;q)=\frac{2}{M \Gamma }S(q)\Lambda _{C}(t;\Gamma)S^\mathsf{T}(q),
\end{equation*}%
where
\begin{equation*}
\begin{split}
\Lambda _{C}(t;\Gamma) = & \mbox{diag}(0, I_{1}(1-\exp (- M \Gamma
t/(2I_{1}))),  \\
& I_{2}(1-\exp (-M \Gamma t/(2I_{2}))), \\
& I_{3}(1- \exp (- M\Gamma t/(2I_{3})))).
\end{split}
\end{equation*}%
If we introduce a $4\times 3$-dimensional matrix $\sigma
(t,q)$ such that
\begin{equation}
\sigma (t;q)\sigma ^{\mathsf{T}}(t;q)=C(t;q),  \label{lbec}
\end{equation}%
e.g., $\sigma (t;q)$ with the columns%
\begin{equation*}
\sigma _{l}(t;q)=\sqrt{\frac{2}{M\Gamma }I_{l}\left(1-{\rm e}^{-\frac{M\Gamma t}{2I_{l}}}\right)}S_{l}q,
\end{equation*}
$l=1,2,3$, then the expression for $\Pi _{I}^{j}(t)$ in (\ref{lbe}) can be written as
\begin{equation*}
\Pi _{I}^{j}(t) = {\rm e}^{-\Gamma J(q)t}\Pi _{I}^{j}(0) +\sqrt{\frac{2M\Gamma }{\beta }}\sum\limits_{l=1}^{3} \sigma_l(t;q) \chi_l^j,
\end{equation*}
where $\chi_l^j$ are independent Gaussian random variables with zero mean and unit variance. We point out a substantial simplification in the calculation of $\sigma$ relative to the equivalent stage of Langevin~B in Ref.~\onlinecite{DHT09}, \trim{made evident by the explicit use of} $S(q)$.

Using the above calculations, \trim{and analogous} procedures for the linear momenta, we obtain the following quasi-symplectic scheme
for (\ref{nl1})--(\ref{nl2}):
\begin{equation}
\begin{split}
\mathbf{P}_{0}& =\mathbf{p},\ \ \mathbf{R}_{0}=\mathbf{r},\ \mathbf{Q}_{0}=%
\mathbf{q},{\ |q^{j}|=1,\ j=1,\ldots ,n} \\
\mathbf{\Pi }_{0}& =\mathbf{\pi },\ \ {\mathbf{q}^{\mathsf{T}}\mathbf{\pi }=0%
} \\
\mathcal{P}_{1,k}& =\mathbf{P}_{k}{\rm e}^{-\gamma h/2}+\sqrt{\frac{m}{\beta }%
(1-{\rm e}^{-\gamma h})}\mathbf{\xi }_{k} \\
\mathit{\Pi }_{1,k}^{j}& = {\rm e}^{-\Gamma J(Q_{k}^{j})\frac{h}{2}}\Pi _{k}^{j}
\\
& {+\sqrt{\frac{4}{\beta }}\sum_{l=1}^{3}\sqrt{I_{l}\left( 1-{\mathrm{e}}^{-%
\frac{M\Gamma h}{4I_{l}}}\right) }S_{l}Q_{k}^{j}\eta _{k}^{j,l},}\
j=1,\ldots ,n,
\end{split}
\label{secla}
\end{equation}%
\begin{equation*}
\begin{split}
\mathcal{P}_{2,k}& =\mathcal{P}_{1,k}+\frac{h}{2}\mathbf{f}(\mathbf{R}_{k},%
\mathbf{Q}_{k}), \\
\mathit{\Pi }_{2,k}^{j}& =\mathit{\Pi }_{1,k}^{j}+\frac{h}{2}F{^{j}}(\mathbf{%
R}_{k},\mathbf{Q}_{k}),\ \ j=1,\ldots ,n, \\
\mathbf{R}_{k+1}& =\mathbf{R}_{k}+\frac{h}{m}\mathcal{P}_{2,k},
\end{split}%
\end{equation*}%
\begin{equation*}
(Q_{k+1}^{j},\mathit{\Pi }_{3,k}^{j})=\Psi _{h}(Q_{k}^{j},\mathit{\Pi }%
_{2,k}^{j}),\ \ j=1,\ldots ,n,
\end{equation*}%
\begin{eqnarray*}
\mathit{\Pi }_{4,k}^{j} &=&\mathit{\Pi }_{3,k}^{j}+\frac{h}{2}F^{j}(\mathbf{R%
}_{k+1},\mathbf{Q}_{k+1}),\ j=1,\ldots ,n, \\
\mathcal{P}_{3,k} &=&\mathcal{P}_{2,k}+\frac{h}{2}\mathbf{f}(\mathbf{R}%
_{k+1},\mathbf{Q}_{k+1}),
\end{eqnarray*}%
\begin{equation*}
\begin{split}
\mathbf{P}_{k+1}& =\mathcal{P}_{3,k}{\rm e}^{-\gamma h/2}+\sqrt{\frac{m}{\beta }%
(1-{\rm e}^{-\gamma h})}\mathbf{\zeta }_{k},\ \  \\
\Pi _{k+1}^{j}& = {\rm e}^{-\Gamma J(Q_{k+1}^{j})\frac{h}{2}}\mathit{\Pi }%
_{4,k}^{j} \\
& {+\sqrt{\frac{4}{\beta }}\sum_{l=1}^{3}\sqrt{I_{l}\left( 1-{\mathrm{e}}^{-%
\frac{M\Gamma h}{4I_{l}}}\right) }S_{l}Q_{k+1}^{j}\varsigma _{k}^{j,l},}\  \\
j& =1,\ldots ,n, \\
k& =0,\ldots ,N-1,
\end{split}%
\end{equation*}%
where $\mathbf{\xi }_{k}=(\xi _{1,k},\ldots ,\xi _{3n,k})^{\mathsf{T}},$ $%
\mathbf{\zeta }_{k}=(\zeta _{1,k},\ldots ,\zeta _{3n,k})^{\mathsf{T}}$ and $%
\eta _{k}^{j}=(\eta _{1,k}^{j},\ldots ,\eta _{3,k}^{j})^{\mathsf{T}},$ $%
\varsigma _{k}^{j}=(\varsigma _{1,k}^{j},\ldots ,\varsigma _{3,k}^{j})^{%
\mathsf{T}},$ $j=1,\ldots ,n,$ with their components being i.i.d.~random
variables with the same law (\ref{n31}).

Properties of the integrator (\ref{secla}),~(\ref{n31}) are summarized in the next
proposition; \trim{we omit the proofs, which are obtainable by standard methods\cite{MT1}}.

\begin{proposition}
\label{prp2}The numerical scheme $(\ref{secla})$,~$(\ref{n31})$ for $(\ref%
{nl1})$--$(\ref{nl2})$ is quasi-symplectic, it preserves $(\ref{a211})$ and 
$(\ref{qpi0})$ and it is of weak order two.
\end{proposition}

We note that in the case of translational degrees of freedom Langevin~B
coincides with the scheme called `OBABO' in Ref.~\onlinecite{BenM13}.

\subsection{Geometric integrator: Langevin C\label{sec:LC}}

This integrator is based on the same splitting (\ref{lb1})--(\ref{lb2}) as
Langevin~B but using a different concatenation: we take half a 
step of a symplectic method for (\ref{lb2}), one step of (\ref{lb1}) using
(\ref{lbe}), and again half a step of (\ref{lb2}). The resulting method
takes the form
\begin{equation}
\begin{split}
\mathbf{P}_{0}& =\mathbf{p},\ \ \mathbf{R}_{0}=\mathbf{r},\ \mathbf{Q}_{0}=%
\mathbf{q}, \\
& |q^{j}|=1,\ j=1,\ldots ,n, \\
\mathbf{\Pi }_{0}& =\mathbf{\pi },\ \ \mathbf{q}^{\mathsf{T}}\mathbf{\pi }=0,
\\
\mathcal{P}_{1,k}& =\mathbf{P}_{k}+\frac{h}{2}\mathbf{f}(\mathbf{R}_{k},%
\mathbf{Q}_{k}), \\
\mathit{\Pi }_{1,k}^{j}& =\Pi _{k}^{j}+\frac{h}{2}F{^{j}}(\mathbf{R}_{k},%
\mathbf{Q}_{k}),\ \ j=1,\ldots ,n, \\
\mathit{R}_{1,k}& =\mathbf{R}_{k}+\frac{h}{2m}\mathcal{P}_{1,k}, \\
(\mathcal{Q}_{1,k}^{j},\mathit{\Pi }_{2,k}^{j})& =\Psi _{h/2}(Q_{k}^{j},%
\mathit{\Pi }_{1,k}^{j}),\ j=1,\ldots ,n, \\
\mathcal{P}_{2,k}& =\mathcal{P}_{1,k}{\rm e}^{-\gamma h}+\sqrt{\frac{m}{\beta }%
(1-{\rm e}^{-2\gamma h})}\mathbf{\xi }_{k} \\
\mathit{\Pi }_{3,k}^{j}& =\mathrm{e}^{-\Gamma J(\mathcal{Q}_{1,k}^{j})h}\mathit{\Pi }%
_{2,k}^{j} \\
& +\sqrt{\frac{4}{\beta }}\sum_{l=1}^{3}\sqrt{I_{l}\left( 1-{\mathrm{e}}^{-%
\frac{M\Gamma h}{2I_{l}}}\right) }S_{l}\mathcal{Q}_{1,k}^{j}\eta
_{k}^{j,l},
\\ & j=1,\ldots ,n,
\end{split}
\label{langC}
\end{equation}
\begin{equation*}
\begin{split}
\mathbf{R}_{k+1}& =\mathit{R}_{1,k}+\frac{h}{2m}\mathcal{P}_{2,k}, \\
(Q_{k+1}^{j},\mathit{\Pi }_{4,k}^{j})& =\Psi _{h/2}(\mathcal{Q}_{1,k}^{j},%
\mathit{\Pi }_{3,k}^{j}),\ \ j=1,\ldots ,n,
\end{split}%
\end{equation*}

\begin{equation*}
\begin{split}
\mathbf{P}_{k+1}& =\mathcal{P}_{2,k}+\frac{h}{2}\mathbf{f}(\mathbf{R}_{k+1},%
\mathbf{Q}_{k+1}), \\
\Pi _{k+1}^{j}& =\mathit{\Pi }_{4,k}^{j}+\frac{h}{2}F{^{j}}(\mathbf{R}_{k+1},%
\mathbf{Q}_{k+1}),\ \ j=1,\ldots ,n,
\end{split}%
\end{equation*}%
where $\mathbf{\xi }_{k}=(\xi _{1,k},\ldots ,\xi _{3n,k})^{\mathsf{T}}$ and $%
\eta _{k}^{j}=(\eta _{1,k}^{j},\ldots ,\eta _{3,k}^{j})^{\mathsf{T}},$
 $j=1,\ldots ,n,$ with their components being i.i.d.~random
variables with the same law (\ref{n31}).

Properties of the integrator (\ref{langC}),~(\ref{n31}) are 
summarized in the next proposition (its proof is omitted here).

\begin{proposition}
\label{prp3}The numerical scheme $(\ref{langC})$,~$(\ref{n31})$ for 
$(\ref{nl1})$--$(\ref{nl2})$ is quasi-symplectic, 
it preserves $(\ref{a211})$ and $(\ref{qpi0})$
and it is of weak order two.
\end{proposition}

We note that in the case of translational degrees of freedom Langevin~C
coincides with the scheme called `BAOAB' in Ref.~\onlinecite{BenM13}, which was shown
there to be the most efficient scheme among various types of splittings of
Langevin equations for systems without rotational degrees of freedom.

\subsection{Numerical scheme for the gradient thermostat\label{sec:Ngt}}

To preserve the length of quaternions in the case of numerical integration
of the gradient system (\ref{a10})--(\ref{a100}), we use ideas of Lie-group
type integrators for deterministic ordinary differential equations on
manifolds (see, e.g., Ref.~\onlinecite{HLW02} and the references therein
and also Ref.~\onlinecite{CasGa95} were such ideas where used for constructing
mean-square approximations for Stratonovich SDEs).  The main idea is to rewrite
the components $Q^{j}$ of the solution to (\ref{a10})--(\ref{a100}) in the form
$Q^{j}(t)=\exp (Y^{j}(t))Q^{j}(0)$ and then solve numerically the SDEs for
the $4\times 4$-matrices $Y^{j}(t)$. To this end, we introduce the $4\times 4$
skew-symmetric matrices:
\begin{eqnarray*}
\mathbb{F}_{j}(\mathbf{r},\mathbf{q})&=&F^{j}(\mathbf{r},\mathbf{q})
q^{j\,\mathsf{T}}-q^{j}(F^{j}(\mathbf{r},\mathbf{q}))^{\mathsf{T}},\\
&&j=1,\ldots ,n.
\end{eqnarray*}%
Note that $\mathbb{F}_{j}(\mathbf{r},\mathbf{q})q^{j}=F^{j}(\mathbf{r},%
\mathbf{q})$ under $|q^{j}|=1$ and the equations (\ref{a100}) can be written
as
\begin{eqnarray}
dQ^{j}&=&\frac{\Upsilon }{M}\mathbb{F}_{j}(\mathbf{R},\mathbf{Q})Q^{j}dt\notag\\
&+&\sqrt{\frac{2\Upsilon}{M\beta}}\sum_{l=1}^{3}S_{l}Q^{j} \star dW_{l}^{j}(t),\notag \\
&&Q^{j}(0)=q^{j},\ \ |q^{j}|=1.  \label{ngt1}
\end{eqnarray}
We also remark that if $\mathbb{F}_{j}(\mathbf{r},\mathbf{q})=0,$ $Q^{j}$
are Wiener processes on the three-dimensional sphere \cite%
{IkWa,Elw,WilRog,GC}. One can show that
\begin{eqnarray*}
Y^{j}(t+h)&=&h\frac{\Upsilon }{M}\mathbb{F}_{j}(\mathbf{R}(t),\mathbf{Q}(t))\\
&+&\sqrt{\frac{2\Upsilon }{M\beta }}\sum_{l=1}^{3}\left(
W_{l}^{j}(t+h)-W_{l}^{j}(t)\right) S_{l}\\&+&\text{terms of higher order.}
\end{eqnarray*}

Consequently, we derived the following numerical method for 
(\ref{a10})--(\ref{a100}):
\begin{align}
\mathbf{R}_{0}& =\mathbf{r},\ \mathbf{Q}_{0}=\mathbf{q},\ |q^{j}|=1,\
j=1,\ldots ,n,  \label{firga} \\
\mathbf{R}_{k+1}& =\mathbf{R}_{k}+h\frac{\upsilon }{m}\mathbf{f}(\mathbf{R}_{k},%
\mathbf{Q}_{k})+\sqrt{h}\sqrt{\frac{2\upsilon }{m\beta }}\mathbf{\xi }_{k},
\notag \\
\mathit{Y}_{k}^{j}& =h\frac{\Upsilon }{M}\mathbb{F}_{j}(\mathbf{R}_{k},%
\mathbf{Q}_{k})+\sqrt{h}\sqrt{\frac{2\Upsilon }{M\beta }}\sum_{l=1}^{3}\eta
_{k}^{j,l}S_{l},  \notag \\
Q_{k+1}^{j}& =\exp (\mathit{Y}_{k}^{j})Q_{k}^{j},\ \ j=1,\ldots ,n,  \notag
\end{align}%
where $\mathbf{\xi }_{k}=(\xi _{1,k},\ldots ,\xi _{3n,k})^{\mathsf{T}}$ and $%
\xi _{i,k}$, $i=1,\ldots ,3n,$ $\eta _{k}^{j,l},$ $l=1,2,3,$ $j=1,\ldots ,n,$
are i.i.d.~random variables with the same law
\begin{equation}
P(\theta =\pm 1)=1/2.  \label{n31simple}
\end{equation}%
Since the matrix $\mathit{Y}_{k}^{j}$ is skew symmetric, the exponent $\exp (%
\mathit{Y}_{k}^{j})$ can be effectively computed using the Rodrigues formula
(see Appendix~\ref{sec:rodr}).

Note that it is sufficient here to use the simpler distribution (\ref{n31simple}) 
than (\ref{n31}) used in the Langevin integrators since the
scheme (\ref{firga}) is of weak order one, while the Langevin integrators
\trim{in this work} are of weak order two (for further reading see,
e.g. Ref.~\onlinecite{MT1}). We proved (the proof is not included here)
that the geometric integrator (\ref{firga})--(\ref{n31simple}) 
possesses properties stated in the following proposition.
\begin{proposition}
\label{prp4}The numerical scheme $(\ref{firga})$--$(\ref{n31simple})$ 
for $(\ref{a10})$--$(\ref{a100})$ preserves the length of quaternions, 
i.e., $|Q_{k}^{j}|=1,\ \ j=1,\ldots ,n\,,$ for all $k$, and it is of weak order one.
\end{proposition}

We note that one can choose $\mathbf{\xi }_{k}$ and $\eta _{k}^{j,l},$ $%
l=1,2,3,$ $j=1,\ldots ,n,$ so that their components are i.i.d.~Gaussian
random variables with zero mean and unit variance. In this case the weak
order of the scheme remains 1st as when we use the simple discrete
distribution (\ref{n31simple}). Let us remark in passing that in the case of
Gaussian random variables the above scheme also converges in the mean-square
sense with order $1/2$. It is not difficult to derive a method of
mean-square order one, which preserves the length of quaternions, but it is not of
applicable interest in our context and hence omitted. As far as we know,
this is the first time when a Lie-group type weak scheme is considered and
applied in the context of stochastic thermostats.

\section{Numerical Experiments\label{sec:num}}
We have implemented Langevin A, B, and C integrators and the weak
1-st order gradient integrator in the simulation of a rigid TIP4P water model
with smoothly truncated electrostatic interactions\cite{RLD10}.
\trim{We simulate a system of 1728 molecules at a density of 989.85\,kg/m$^{3}$, with periodic boundary conditions.}
All simulation runs start from the same initial state, which is a well-equilibrated liquid state at 300\,K obtained
in a long simulation with $h=1$\,fs.
After a further 20\,000-step equilibration run (which gives a sufficiently long equilibration time for all the
considered time steps, taking into account that the initial state is already well equilibrated), the measurements are accumulated during the subsequent 200\,000 steps.

During the simulations, we monitored the preservation of
the constraint $|q^j| = 1$ by all integrators and $q^{j\,\mathsf{T}} \pi^j = 0$ by
the Langevin integrators, $j = 1, \ldots, n$. \trim{The behavior indicates that these constraints hold to within the round-off
error, which gradually accumulates with the number of steps, but is
independent of the step size $h$.}


\begin{figure}[tbp]
\begin{center}
\includegraphics[width=8.0cm]{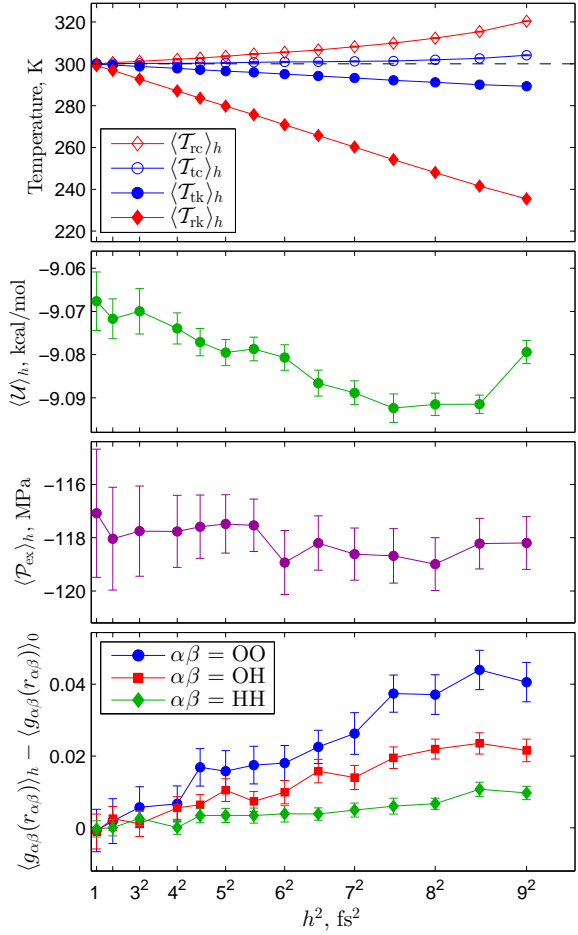}
\end{center}
\caption{Langevin A thermostat with $\protect\gamma =5\,$ps$^{-1}$ and
$\Gamma =10\,$ps$^{-1}$. Error bars denote estimated 95\% confidence intervals
in the measured quantities. The bottom plot illustrates numerical integration error in the evaluation of the RDFs near the first maximum.  The estimated values of $\langle g_{\alpha\beta}(r_{\alpha\beta}) \rangle_0$ for $\alpha\beta = $ OO, OH, and HH are
$3.007(4)$, $1.490(3)$, and $1.283(2)$, respectively.}
\label{fig:LanA}
\end{figure}

\begin{figure}[tbp]
\begin{center}
\includegraphics[width=8.0cm]{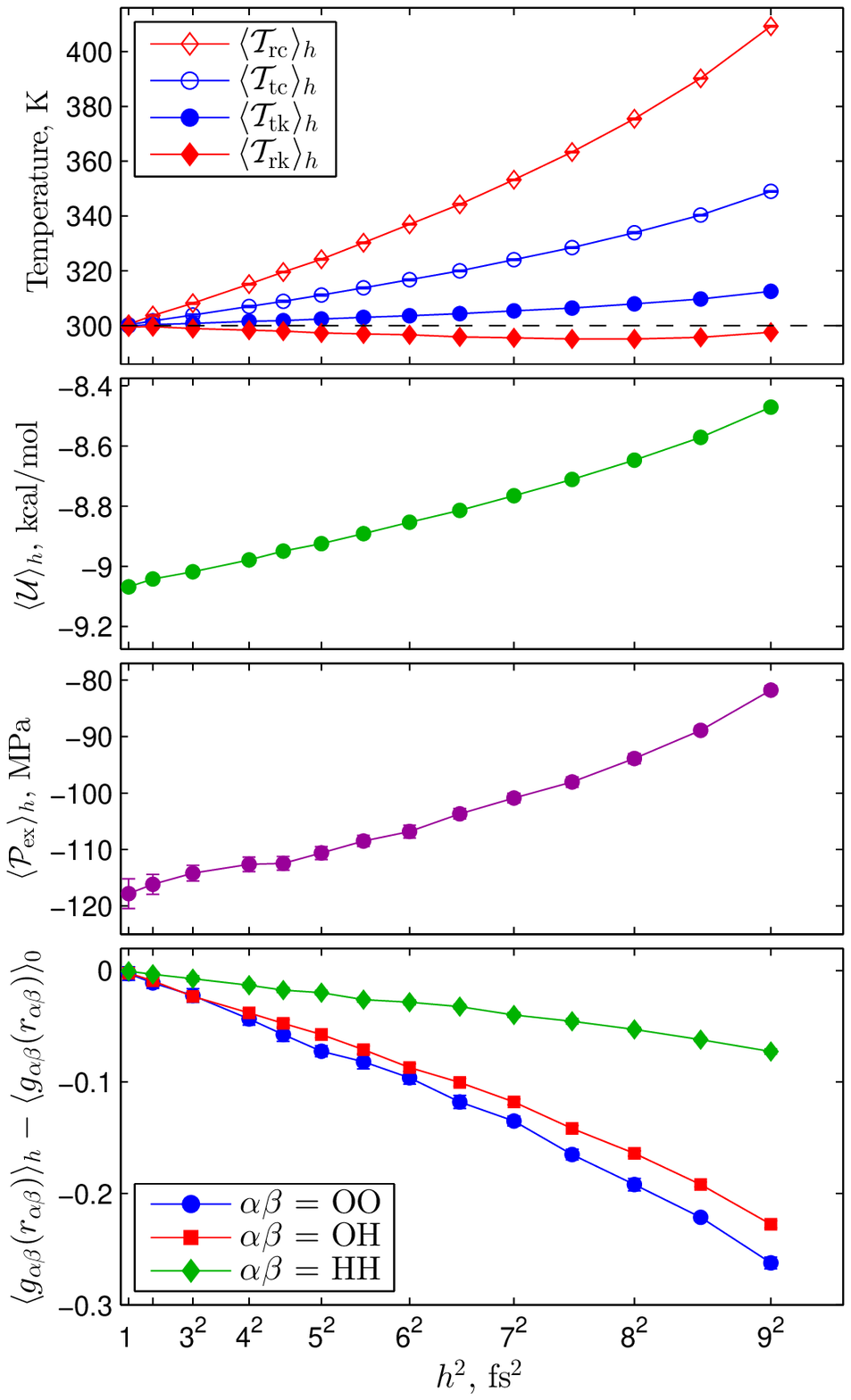}
\end{center}
\caption{Langevin B thermostat with $\protect\gamma =5\,$ps$^{-1}$ and
$\Gamma =10\,$ps$^{-1}$. Error bars denote estimated 95\% confidence intervals
in the measured quantities. The bottom plot illustrates numerical integration error in the evaluation of the RDFs near the first maximum.  The estimated values of $\langle g_{\alpha\beta}(r_{\alpha\beta}) \rangle_0$ for $\alpha\beta = $ OO, OH, and HH are
$3.009(4)$, $1.492(2)$, and $1.284(2)$, respectively.}
\label{fig:LanB}
\end{figure}

\begin{figure}[tbp]
\begin{center}
\includegraphics[width=8.0cm]{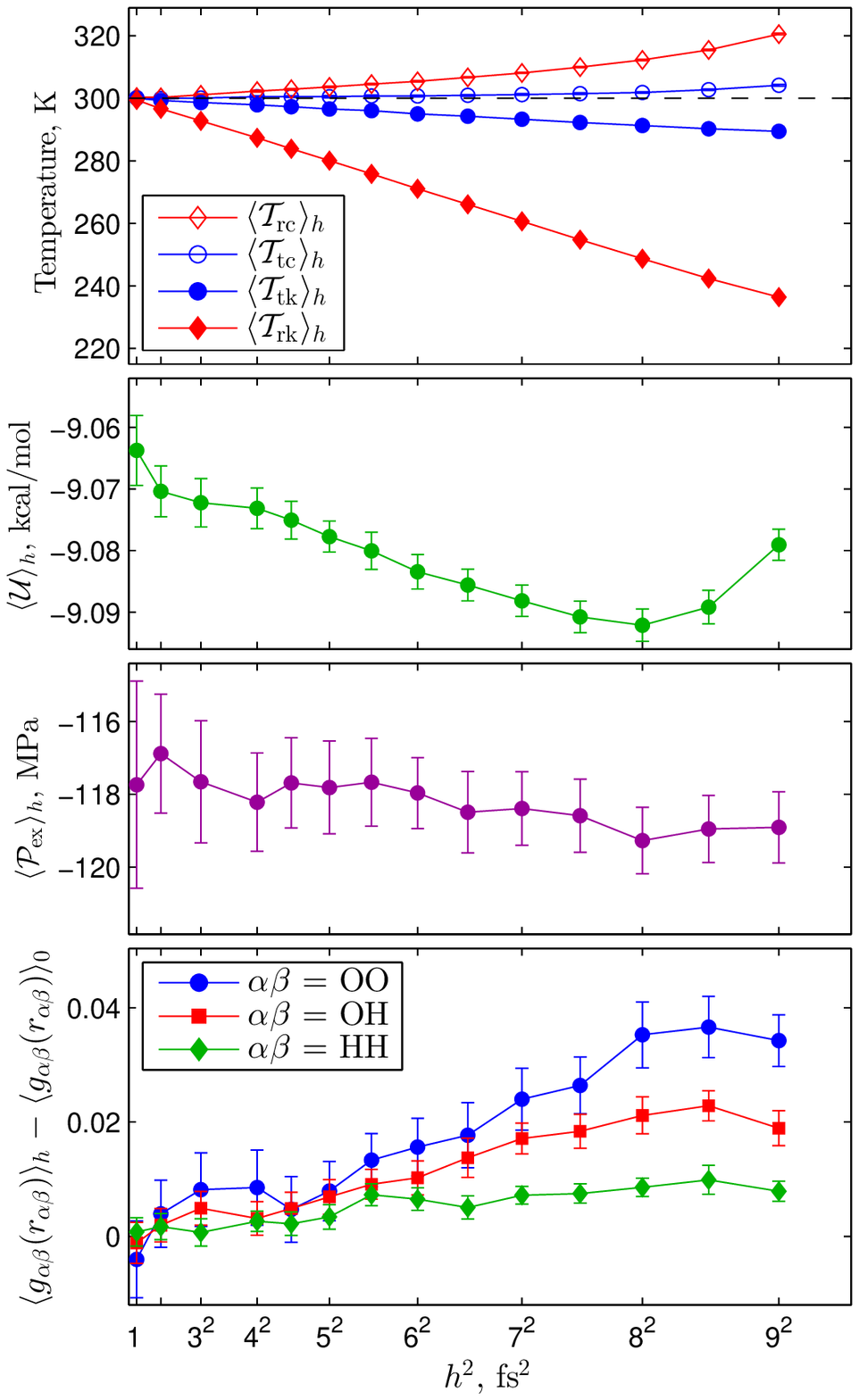}
\end{center}
\caption{Langevin C thermostat with $\protect\gamma =5\,$ps$^{-1}$ and
$\Gamma =10\,$ps$^{-1}$. Error bars denote estimated 95\% confidence intervals
in the measured quantities. The bottom plot illustrates numerical integration error in the evaluation of the RDFs near the first maximum.  The estimated values of $\langle g_{\alpha\beta}(r_{\alpha\beta}) \rangle_0$ for $\alpha\beta = $ OO, OH, and HH are
$3.009(4)$, $1.490(2)$, and $1.282(2)$, respectively.}
\label{fig:LanC}
\end{figure}

\begin{figure}[tbp]
\begin{center}
\includegraphics[width=8.0cm]{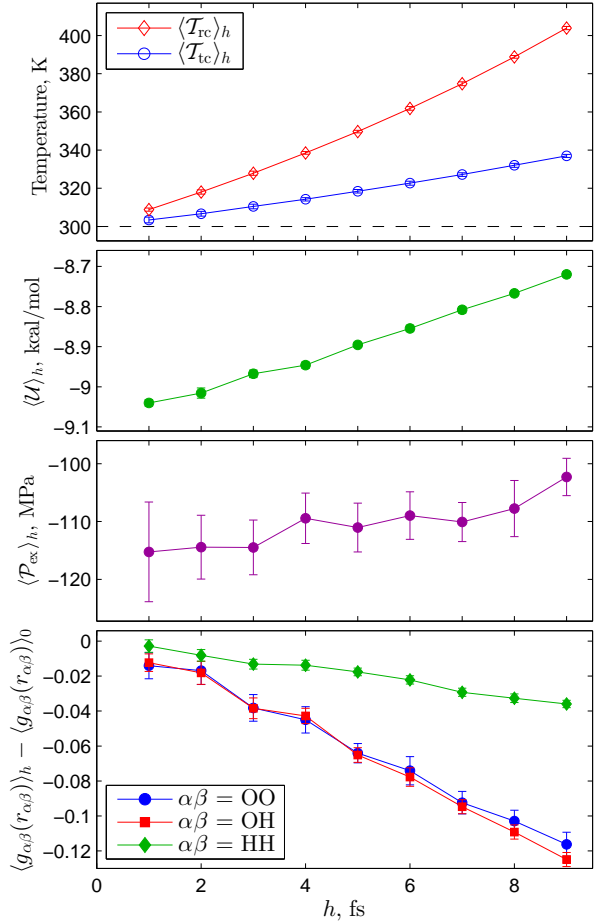}
\end{center}
\caption{Gradient thermostat with $\protect\upsilon =4\,$fs and
$\Upsilon = 1\,$fs. Error bars denote estimated 95\% confidence intervals
in the measured quantities. The bottom plot illustrates numerical integration error in the evaluation of the RDFs near the first maximum.  The estimated values of $\langle g_{\alpha\beta}(r_{\alpha\beta}) \rangle_0$ for $\alpha\beta = $ OO, OH, and HH are
$3.012(9)$, $1.491(7)$, and $1.284(4)$, respectively.}
\label{fig:Grad}
\end{figure}

The following quantities are measured during simulation runs:
\begin{itemize}
\item Translational kinetic temperature
\[ \langle \mathcal{T}_\mathrm{tk}\rangle_h  =
\frac{\langle \mathbf{p}^{\mathsf{T}}\mathbf{p}\rangle_h }{3mk_B n};
\]
\item Rotational kinetic temperature
\[ \langle \mathcal{T}_\mathrm{rk}\rangle_h  =
\frac{2\left\langle\sum_{j=1}^{n}\sum_{l=1}^{3}V_{l}(q^{j},\pi ^{j})\right\rangle_h}{3k_B n};
\]
\item Translational configurational temperature
\[\langle \mathcal{T}_\mathrm{tc}\rangle_h  =
\frac{\left\langle\sum_{j=1}^{n} | \nabla_{r^j} U |^2 \right\rangle_h}
   {k_B\left\langle\sum_{j=1}^{n} \nabla_{r^j}^2 U \right\rangle_h};
\]
\item Rotational configurational temperature\cite{Chialvo01}
\[\langle \mathcal{T}_\mathrm{rc}\rangle_h  =
\frac{\left\langle\sum_{j=1}^{n} | \nabla_{\omega^j} U |^2 \right\rangle_h}
   {k_B\left\langle\sum_{j=1}^{n} \nabla_{\omega^j}^2 U \right\rangle_h},
\]
where $\nabla_{\omega^j}$ is the angular gradient operator for molecule $j$;
\item Potential energy per molecule
\[ \langle \mathcal{U} \rangle_h = \frac{1}{n} \langle U \rangle_h;
\]
\item Excess pressure
\[\langle \mathcal{P}_\mathrm{ex} \rangle_h =
-\frac{\left\langle\sum_{j=1}^{n} r^{j\,\mathsf{T}} f^j\right\rangle_h}{3V},
\]
where $V$ is the system volume;
\Rf{
\item Radial distribution functions (RDFs) between oxygen (O) and hydrogen (H) interaction sites
\[ \langle g_{\alpha\beta}(r) \rangle_h\,,\]
where $\alpha\beta =$ OO, OH, and HH.
}
\end{itemize}
Angle brackets with subscript $h$ represent the average over a simulation run with time step $h$.

We run simulations with different time steps $h$ ranging from 1 to 9\,fs.
\trim{Due to the discretization errors, measured quantities depend on $h$; for an
integrator of order $p$ estimating a quantity $A$\cite{TAT90,MT1}:
\begin{equation}
  \langle A \rangle_h = \langle A \rangle_0 + E_A h^p + \mathcal{O}(h^{p+1}),
\label{eq:error-h}
\end{equation}}
\Rf{For the RDFs, the dependence on $h$ is observed at every value of $r$, that is
\[\langle g_{\alpha\beta}(r) \rangle_h = \langle g_{\alpha\beta}(r) \rangle_0
+ E_{g_{\alpha\beta}}(r)h^p + \mathcal{O}(h^{p+1})\,, \]
(cf. Ref.~\onlinecite{RLD10}).
}

\begin{table*}[tb]
\caption{\label{tab:results} Results for Langevin A, B, and C thermostats with $\gamma = 5\,$ps$^{-1}$ and $\Gamma = 10\,$ps$^{-1}$ and gradient thermostat with $\upsilon = 4\,$fs and $\Upsilon = 1\,$fs.  Values of $\langle A \rangle_0$ and $E_A$, defined in Eq.~(\ref{eq:error-h}), were obtained by linear regression from $\langle A \rangle_h$ for $h \leq 6\,$fs for Langevin integrators and for $h \leq 4\,$fs for the gradient integrator.  Quantities $E_A$ are measured in the units of the corresponding quantity $A$ per fs$^p$, where $p = 2$ for Langevin integrators and $p = 1$ for the gradient integrator.  \Rf{The last three rows present values of the RDFs near the first maximum: $r_\mathrm{OO} = 2.78\,$\AA, $r_\mathrm{OH} = 1.82\,$\AA, and $r_\mathrm{HH} = 2.38\,$\AA.} Numbers in parentheses indicate the statistical error in the last digit(s) shown with a 95\% confidence level.}
\begin{ruledtabular}
\begin{tabular}{ccccccccc}
 &\multicolumn{2}{c}{Langevin A} & \multicolumn{2}{c}{Langevin B} & \multicolumn{2}{c}{Langevin C} & \multicolumn{2}{c}{Gradient}\\
 $A$, unit & $\langle A \rangle_0$ & $E_A$ & $\langle A \rangle_0$ & $E_A$ & $\langle A \rangle_0$ & $E_A$ & $\langle A \rangle_0$ & $E_A$ \\ \hline
 $\mathcal{T}_\mathrm{tk}$, K & $300.0(2)$    & $-0.136(8)$ & $299.9(2)$    & $0.100(13)$  & $300.0(2)$ & $-0.135(7)$ & $-$ & $-$   \\
 $\mathcal{T}_\mathrm{rk}$, K & $299.9(2)$    & $-0.808(8)$ & $299.8(3)$    & $-0.092(13)$ & $300.1(2)$ & $-0.803(8)$ & $-$ & $-$   \\
 $\mathcal{T}_\mathrm{tc}$, K & $300.1(3)$    & $0.022(13)$ & $299.9(4)$    & $0.45(2)$    & $300.1(3)$ & $0.021(13)$ & $299.6(1.0)$ & $3.6(5)$   \\
 $\mathcal{T}_\mathrm{rc}$, K & $299.8(3)$    & $0.158(11)$ & $299.6(4)$    & $0.99(2)$    & $299.9(3)$ & $0.152(11)$ & $298.6(1.6)$ & $9.9(4)$   \\
 $\mathcal{U}$, kcal/mol      & $-9.068(4)$   & $-0.0004(2)$& $-9.071(4)$   & $0.0059(2)$  & $-9.066(3)$ & $-0.0005(2)$ & $-9.075(11)$ & $0.033(4)$   \\
 $\mathcal{P}_\mathrm{ex}$, MPa & $-117.4(1.3)$ & $-0.02(5)$  & $-117.4(1.6)$ & $0.27(9)$   & $-117.5(1.4)$ & $-0.01(5)$ & $-118(11)$ & $1.7(2.8)$   \\
 $g_\mathrm{OO}(r_\mathrm{OO})$ & $3.007(4)$ & $0.0006(2)$ & $3.009(4)$ & $-0.0027(2)$ & $3.009(4)$ & $0.0004(2)$ & $3.012(9)$ & $-0.011(4)$ \\
 $g_\mathrm{OH}(r_\mathrm{OH})$ & $1.490(3)$ & $0.0003(2)$&$1.492(2)$ & $-0.0024(2)$ & $1.490(2)$& $0.00028(9)$& $1.491(7)$ & $-0.011(2)$ \\
 $g_\mathrm{HH}(r_\mathrm{HH})$ & $1.283(2)$ &$0.00012(7)$&$1.284(2)$ & $-0.00082(6)$ &$1.282(2)$& $0.00018(7)$& $1.284(4)$ & $-0.004(2)$
\end{tabular}
\end{ruledtabular}
\end{table*}

The results are presented in Figs.~\ref{fig:LanA}-\ref{fig:Grad}.
\Rf{For the RDFs, we plot the difference
$\langle g_{\alpha\beta}(r) \rangle_h - \langle g_{\alpha\beta}(r) \rangle_0$
evaluated at $r = r_{\alpha\beta}$ near the first maximum of the corresponding RDF,
where the magnitude of the numerical error is the largest.}
Error bars represent 95\% confidence intervals estimated using the
block averaging approach\cite{FrenkelBook}.
As expected \trim{$\langle A \rangle_h$ depends linearly on $h^2$ for
the Langevin integrators and on $h$ for the gradient integrator at small $h$}.
Using linear regression, we calculate the values of $\langle A \rangle_0$
and $E_A$ for the measured quantities as defined in Eq.~(\ref{eq:error-h}).
The results are presented in Table~\ref{tab:results}.  We observe that for all
numerical methods all four measures of the system temperature converge in the limit $h \to 0$
to the correct value of $T = 300\,$K set by the thermostats.
At the same time, the leading discretization error terms, characterised by $E_A$,
are different for different measures of temperature and for different
numerical methods.  We also see that the estimated values of $\langle \mathcal{U} \rangle_0$, $\langle \mathcal{P}_\mathrm{ex} \rangle_0$, \Rf{and $\langle g_{\alpha\beta}(r) \rangle_0$} agree for all integrators, which is an indicator
that all the integrators sample from the same ensemble.

While here we present only results for one set of thermostat parameters
($\gamma =5\,$ps$^{-1}$ and $\Gamma =10\,$ps$^{-1}$ for the Langevin thermostat and
$\protect\upsilon =4\,$fs and $\Upsilon = 1\,$fs for the gradient thermostat),
we have \trim{tested} a wide range of
parameter values.  The results obtained with Langevin A and B integrators are
identical within the sampling errors to those of the corresponding integrators
in Refs.~\onlinecite{DHT09,RLD10}.  This is expected since, as demonstrated in Appendix~\ref{sec:just},
the components of $\pi^j$ parallel to $q^j$ do not influence the physical properties of the system.

\trim{Due to} the exact treatment of the Ornstein-Uhlenbeck process, Langevin B and C integrators can be used with arbitrarily large values of $\gamma$ and $\Gamma$.  \trim{Langevin A} breaks down for $\gamma$ larger than about 100\,ps$^{-1}$ and for $\Gamma$ larger than about 200\,ps$^{-1}$ (see also Fig.~9 in Ref.~\onlinecite{DHT09}).  \trim{In terms of sampling efficiency, the optimal damping parameters for this system are $\gamma = 3$-$6\,$ps$^{-1}$ and $\Gamma = 7$-$15\,$ps$^{-1}$.  Larger values of $\gamma$ and $\Gamma$ slow system evolution and hence inhibit sampling; smaller values make the thermostat weaker and increase the time required to reach equilibrium.} When the system is already well equilibrated, it is reasonable to use smaller values
of $\gamma$ and $\Gamma$, especially if one wants to measure time-dependent properties such as time autocorrelation functions. We also note that stronger coupling to the thermostat stabilizes the integrator:  it is unstable at around $h = 8\,$fs for very small values of $\gamma$ and $\Gamma$, but is stable up to about $h = 10\,$fs when $\gamma + \Gamma$ is larger than about 8\,ps$^{-1}$.

With the gradient system, scaling $\upsilon$ and $\Upsilon$ together does not change the properties of the trajectories, only the speed at which these trajectories are traversed.  This property is manifest in the integrator by the fact that thermostat parameters appear together with the time step $h$ as $h\upsilon$ and $h\Upsilon$.
\trim{Consequently}, it is always possible to use larger $h$ with smaller $\upsilon$ and $\Upsilon$; clearly, however, this does not represent better sampling. For simplicity of numerical experimentation, we use the same range of $h$ values as with the Langevin integrators, while adjusting $\upsilon$ and $\Upsilon$ to achieve optimal simulation efficiency.  For a given $h$, it is preferable to use larger values of $\upsilon$ and $\Upsilon$ \trim{and thereby
make larger effective integration steps.  However, the gradient integrator becomes unstable when $h\upsilon$ exceeds approximately $150\,$fs$^2$ for small $h\Upsilon$ and $h\Upsilon$ exceeds approximately $70\,$fs$^2$ for small $h\upsilon$}.  Also, the linear dependence of the discretization errors on $h$ in Eq.~(\ref{eq:error-h}) extends only up to $h\upsilon \approx 20\,$fs$^2$ for small $h\Upsilon$ and up to $h\Upsilon \approx 5\,$fs$^2$ for small $h\upsilon$.  In Fig.~\ref{fig:Grad} we present results for the gradient integrator with $\upsilon =4\,$fs and $\Upsilon = 1\,$fs and use simulation runs with $h \leq 4\,$fs to obtain linear regression results presented in Table~\ref{tab:results}.

\Rf{It is also possible to construct deterministic thermostats (e.g.~Nos\'e-Hoover, Nos\'e-Poincar\'e)
for the integrator in Ref.\,\onlinecite{qua02}.  A comprehensive study of such thermostats and their comparison to
the Langevin integrators of Ref.\,\onlinecite{DHT09} has been presented in Ref.\,\onlinecite{RLD10}, demonstrating the efficacy of the Langevin approach.
As the behavior of the new integrators Langevin A, B and C is very similar to that of the old Langevin A, B and A from Ref.\,\onlinecite{RLD10}, respectively, we do not repeat the comparison explicitly here.}

\section{Discussion and Conclusions}\label{sec:concl}

\trim{When simulating, we are interested in accuracy and computational cost. The cost of a thermostat + integrator is usually dominated by the number of force calculations per step; since all four integrators presented here use one force calculation per step, the computational costs for a given number of steps are essentially equal.} \trim{Accuracy is determined by the size of the errors. The important sources of errors that vary between our methods are: (1) random statistical error due to incomplete sampling of the equilibrium ensemble; and (2) systematic numerical integration (discretization) error due to approximating system evolution with a finite time step\cite{MT7,MST10,Johan10}}.

\trim{The numerical integration error can be reduced by decreasing the step size $h$, at the cost of increased sampling error unless the number of steps (and hence overall cost) grow}.  As we have seen in our numerical experiments, the size of the numerical error in the measurement of a quantity $A$ is also determined by the value of $E_A$ [see Eq.~(\ref{eq:error-h})], which is influenced by the choice of a thermostat and its parameters, such as damping coefficients.  Therefore, a better way to reduce the numerical error is to choose a thermostat and its parameters so that the magnitude of $E_A$ is small for the quantity of interest $A$.  In some cases it is possible to make $E_A = 0$, which means that for the measurement of $A$ the numerical integrator behaves as a method of higher order~\cite{RLD10,BenM13,LMT14}.  Since it is not always possible to make $E_A$ small simultaneously for all $A$, when choosing the thermostat and its parameters we have to consider which quantity we need to determine most accurately. 

Comparing the performance of the three Langevin integrators, we observe \trim{in the experiments from Section~\ref{sec:num}} that Langevin B is better at controlling the kinetic temperatures, while Langevin A and C are better for more accurate measurements of the configurational temperatures, potential energy, excess pressure, \Rf{and radial distribution functions}.
\Rs{Similar relative performance of different thermostats has been noted elsewhere\cite{RLD10,BenM13,Referee2b}. Since in most cases the quantities of interest are configurational (i.e. they depend on particle positions), Langevin A and C should be the integrators of choice.} \Rf{It is also worth noting that for Langevin A and C, potential energy, excess pressure and radial distribution functions are all within 2\% of the true values even at $h=9$\,fs, when the majority of the temperature measures show significantly larger deviations. However, in every particular case, it is important to estimate numerical errors for quantities of interest in order to determine the appropriate time step for the simulations. }

\trim{In addition, Langevin C  can be used with arbitrarily large values of $\gamma$ and $\Gamma$, which could be desirable if the thermostat is needed to also play the role of an implicit solvent. It is interesting to note in passing that, despite the very different structure of Langevin A and C, their numerical errors for various measured quantities are surprisingly similar.  We do not have an explanation for this ``coincidence'', and further analysis of the integrators is required to understand it.}

\trim{Comparing the Langevin integrators and the gradient integrator, the statistical errors in the results (as measured by the estimated error bars on measurements) of the gradient method are much larger. The sampling efficiency of the gradient method is therefore much lower}. Supporting this observation, the mean-square displacement of molecules after the same number of integration steps is about 10 times larger for the Langevin integrators than for the gradient integrator with the same step size $h$ and maximum possible values of
$\upsilon$ and $\Upsilon$.  Thus the gradient integrator \trim{obtains a factor of 10 fewer decorrelated samples within a given time.}

\trim{Overall, we find that the Langevin A and C integrators perform best. Even though we reached this conclusion} based on the observations of a particular molecular system, we believe they are sufficiently generic.  At the same time, there might be situations when the Brownian thermostat introduced in this paper is preferable to the Langevin one.  We also note that the Brownian thermostat integrator is of order one while the Langevin integrators are of order two. Further work in developing numerical methods for gradient systems is needed.

\section*{Acknowledgment}
This work was partially supported by the Computer Simulation of Condensed Phases (CCP5) Collaboration Grant, which is part of the EPSRC grant EP/J010480/1. T.E.O. also acknowledges funding from University College, Oxford.  \our{This research used the ALICE High Performance Computing Facility at the University of Leicester.}

\appendix{}
\section{Evaluation of the rotational force\label{sec:forces}}
To calculate the rotational force, $F^{j}(%
\mathbf{r},\mathbf{q})=-\tilde{\nabla}_{q^{j}} U(\mathbf{r},\mathbf{q})\in
T_{q^{j}}\mathbb{S}^{3},$ it is necessary to evaluate a directional derivative of the potential tangent to the
three dimensional sphere $\mathbb{S}^{3}$, $\tilde{\nabla}_{q^{j}} U(\mathbf{r},\mathbf{q})$. If the potential is expressed in quaternions, it is natural to calculate this derivative directly via
\begin{equation*}
\tilde{\nabla}_{q^{j}} U(\mathbf{r},\mathbf{q}) = {\nabla}_{q^{j}} U(\mathbf{r},\mathbf{q}) - (q^{j\,\mathsf{T}} {\nabla}_{q^{j}} U(\mathbf{r},\mathbf{q})) q^j,
\end{equation*}
where ${\nabla}_{q^{j}}$ is a conventional 4-component gradient.

Alternatively, the rotational force $F^{j}(\mathbf{r},\mathbf{q})$ can be represented as $%
F^{j}(\mathbf{r},\mathbf{q})=2S(q^{j})(0,\tau ^{j\,\mathsf{T}})^{\mathsf{T%
}}$, where $\tau ^{j}=(\tau _{1}^{j},\tau _{2}^{j},\tau _{3}^{j})^{\mathsf{T}%
}\in \mathbb{R}^{3}$ is the torque acting on molecule $j$ in the body-fixed
reference frame~\cite{qua02}. To illustrate how $F^{j}(\mathbf{r},\mathbf{q}%
) $ can be computed, we consider a specific example -- a system of rigid
molecules with pairwise interaction between interaction sites within
molecules:
\begin{equation}
U(\mathbf{r},\mathbf{q})=\sum_{j}\sum_{m<j}\sum_{\alpha ,\beta }u_{\alpha
\beta }(|r^{j,\alpha }-r^{m,\beta }|)\,,  \label{eq:pairwise}
\end{equation}%
where $r^{j,\alpha }=r^{j}+\mathbf{A}^{\mathsf{T}}(q^{j})d^{\alpha }$ is the
coordinate of the interaction site $\alpha $ within molecule $j$, with $%
d^{\alpha }$ being the site coordinate relative to the center of mass of a
molecule in the body-fixed reference frame. Here
\begin{equation}
\mathbf{A}(q)=2\left[
\begin{array}{ccc}
q_{0}^{2}+q_{1}^{2}-\frac{1}{2} & q_{1}q_{2}+q_{0}q_{3} &
q_{1}q_{3}-q_{0}q_{2} \\
q_{1}q_{2}-q_{0}q_{3} & q_{0}^{2}+q_{2}^{2}-\frac{1}{2} &
q_{2}q_{3}+q_{0}q_{1} \\
q_{1}q_{3}+q_{0}q_{2} & q_{2}q_{3}-q_{0}q_{1} & q_{0}^{2}+q_{3}^{2}-\frac{1}{%
2}%
\end{array}%
\right]  \label{eq:rotmat}
\end{equation}%
is the rotational matrix expressed in terms of quaternion coordinates. Force
acting on the interaction site $\alpha$ of molecule $j$ is given by
\begin{eqnarray}
f^{j,\alpha }&=&-\nabla _{r^{j,\alpha }}U(\mathbf{r},\mathbf{q}) \notag \\
&=&-\sum_{m\neq j}\sum_{\beta }u_{\alpha \beta }^{\prime }(|r^{j,\alpha }-r^{m,\beta }|)%
\frac{r^{j,\alpha }-r^{m,\beta }}{|r^{j,\alpha }-r^{m,\beta }|}. \notag \\
\end{eqnarray}
The total force acting on molecule $j$ is
$f^{j}=\sum_{\alpha }f^{j,\alpha }$.  The torque acting on molecule $j$
in the body-fixed reference frame is given by
\begin{equation}
\tau ^{j}=\sum_{\alpha }d^{\alpha }\times (\mathbf{A}(q^{j})f^{j,\alpha })\,.
\end{equation}%

\trim{
\section{The Langevin thermostat of Ref.~\onlinecite{DHT09} does not preserve $q^{j\,\mathsf{T}}\pi ^{j}=0$}
\label{app:old}
The Langevin thermostat for rigid body dynamics of
Ref.~\onlinecite{DHT09} is given by (\ref{nl1}) and:
\begin{eqnarray}
dQ^{j} &=&\frac{1}{4}S(Q^{j})DS^\mathsf{T}(Q^{j})\Pi ^{j}dt,\
Q^{j}(0)=q^{j}, \label{lt2} \\ && |q^{j}|=1, \notag \\
d\Pi ^{j} &=&\frac{1}{4}\sum_{l=1}^{3}\frac{1}{I_{l}}
\left(\Pi^{j\,\mathsf{T}}S_{l}Q^{j}\right) S_{l}\Pi ^{j}dt +{F^{j}}(\mathbf{R},\mathbf{Q})dt \notag
 \\
&& -\Gamma J(Q^{j})\Pi ^{j}dt+\sqrt{\frac{2M\Gamma }{\beta }}dW^{j}(t), \notag \\
&& \Pi ^{j}(0)=\pi ^{j},\ \ \ j = 1,\ldots ,n,  \notag
\end{eqnarray}
where $(\mathbf{w}^{\mathsf{T}},\mathbf{W}^{\mathsf{T}})^{\mathsf{T}}=(w^{1\,%
\mathsf{T}},\ldots ,w^{n\,\mathsf{T}},W^{1\,{\mathsf{T}}},\ldots ,$ $W^{n\,%
\mathsf{T}})^{\mathsf{T}}$ is a $(3n+4n)$-dimensional standard Wiener
process with $w^{j}=(w_{1}^{j},w_{2}^{j},w_{3}^{j})^{\mathsf{T}}$ and $%
W^{j}=(W_{0}^{j},W_{1}^{j},W_{2}^{j},W_{3}^{j})^{\mathsf{T}}$; 
all other notation is identical to (\ref{nl1})--(\ref{nl2}).

This thermostat  preserves the quaternion length (\ref{a211}).
At the same time, the condition $q^{j\,\mathsf{T}}\pi ^{j}=0$ is not preserved. Let us introduce $\Omega ^{j}(t):=\half(Q^{j}(t))^{\mathsf{T}}\Pi ^{j}(t)$.  We see that $2\Omega^{j}(t)$ represents the component of the angular momenta $\Pi^{j}(t)$ parallel to the rotational coordinates $Q^{j}(t)$.
Recall that $Q^{j}(t)$, being unit quaternions, are constrained to unit spheres; therefore, as in the deterministic case, the quantities $\Omega ^{j}(t)$ should be zero from the physical point of view, i.e., $\Pi^{j}(t)\in T_{Q^{j}}\mathbb{S}^{3}$.
However, the Langevin thermostat of Ref.~\onlinecite{DHT09} does not keep $\Pi
^{j}(t)$ on the tangent space $T_{Q^{j}}\mathbb{S}^{3}$.  Indeed, by direct
calculations we obtain
\begin{equation}
d\Omega^{j}=\half\sqrt{2M\Gamma /\beta }Q^{j\,\mathsf{T}}dW^{j}(t),\
j=1,\ldots ,n\,.  \label{YY}
\end{equation}%
Consequently, if $\Omega^{j}(0)=0$ then $E\Omega^{j}(t)=0$ and $E\left[
\Omega^{j}(t)\right]^{2}=M\Gamma t/(2\beta)$.
}

\section{Proof that the Langevin thermostat from Ref.~\protect\onlinecite{DHT09} is
correct for physical quantities\label{sec:just}}
In this appendix we show that $\Omega^{j}(t)$ being non-zero has no consequences for the measurement of
physical quantities.  First, we modify the solution $X(t)$ of the SDEs 
(\ref{nl1}),~(\ref{lt2}) to turn it into an ergodic process. Let us introduce
\begin{eqnarray*}
\tilde{\Pi}^{j}(t)&:=&\Pi^{j}(t)-(Q^{j}(t))^{\mathsf{T}}\Pi
^{j}(t)Q^{j}(t)\\ &=&\Pi^{j}(t)-Y^{j}(t)Q^{j}(t),\ \ j=1,\ldots ,n,
\end{eqnarray*}%
which are the projections of $\Pi^{j}(t)$ on the tangent space $T_{Q^{j}}%
\mathbb{S}^{3},$ i.e., $\tilde{\Pi}^{j}$ are orthogonal to $Q^{j}$.  By
elementary calculations, we get
\begin{eqnarray}
dQ^{j} &=&\frac{1}{4}S(Q^{j})DS^\mathsf{T}(Q^{j})\tilde{\Pi}^{j}dt, \label{tilda_pi} \\
&& Q^{j}(0)=q^{j},\ |q^{j}|=1, \hspace*{2.3cm}{\ } \notag
\end{eqnarray}
\begin{eqnarray*}
d\tilde{\Pi}^{j} &=&\frac{1}{4}\sum_{l=1}^{3}\frac{1}{I_{l}}\left( \tilde{\Pi}
^{j\,\mathsf{T}}S_{l}Q^{j}\right) S_{l}\tilde{\Pi}^{j}dt\notag\\
&+&{F^{j}}(\mathbf{R},\mathbf{Q})dt-\Gamma J(Q^{j})\tilde{\Pi}^{j}dt\notag \\
&+&\sqrt{\frac{2M\Gamma }{\beta }}%
(I-Q^{j}Q^{j\,\mathsf{T}})dW^{j}(t),  \notag \\
&&\tilde{\Pi}^{j}(0) = \pi^{j},\ {q^{j\,\mathsf{T}}\pi^{j}=0,}\ j=1,\ldots ,n.
\notag
\end{eqnarray*}%
We note the fact that the solution to (\ref{nl1}),~(\ref{tilda_pi}) 
preserves the constraint $(Q^{j}(t))^{\mathsf{T}}\tilde{\Pi}^{j}(t)=0$ 
in addition to the constraint (\ref{a211}), i.e., unlike $\Pi^{j}(t)$, $\tilde{\Pi}^{j}(t)$
have an appropriately constrained dynamics from the physical point of view.

Let $\varphi(x):~\mathbb{D\rightarrow R}$ be a function with polynomial
growth at infinity.  Recall that
\begin{eqnarray*}
\mathbb{D}&=&\{x=(\mathbf{r}^{\mathsf{T}},\mathbf{p}^{\mathsf{T}},\mathbf{q}^{%
\mathsf{T}},\bm{\pi}^{\mathsf{T}})^{\mathsf{T}}\in \mathbb{R}^{14n}:\\
&&|q^{j}|=1,(q^{j})^{\mathsf{T}}\pi ^{j}=0,\ \ j=1,\ldots ,n\}.
\end{eqnarray*}%
Assume that the process $\tilde{X}(t):=(\mathbf{R}^{%
\mathsf{T}}(t),\mathbf{P}^{\mathsf{T}}(t),\mathbf{Q}^{\mathsf{T}}(t),\mathbf{%
\tilde{\Pi}}^{\mathsf{T}}(t))^{\mathsf{T}}$ with $\mathbf{\tilde{\Pi}}=(
\tilde{\Pi}^{1\,\mathsf{T}},\ldots ,\tilde{\Pi}^{n\,\mathsf{T}})^{\mathsf{T}}$
is ergodic, i.e., there exists a unique invariant measure $\mu$ of $\tilde{X}$
and independently of $x\in \mathbb{D}$ there exists the limit
(see, for example, Refs.~\onlinecite{Has,Soize} and references therein):%
\begin{equation}
\lim_{t\rightarrow \infty }E\varphi (\tilde{X}(t;x_{0}))=\int_{\mathbb{D}%
}\varphi (x)\,d\mu (x):=\varphi ^{erg},  \label{PA31}
\end{equation}%
where $\tilde{X}(t;x_{0})$ is attributed to the solution $\tilde{X}(t)$ of 
(\ref{nl1}), (\ref{tilda_pi}) with the initial condition $\tilde{X}(0)=\tilde{%
X}(0;x_{0})=x_{0}\in \mathbb{D}$. Using the stationary Fokker-Planck
equation, it is not difficult to check that $\mu$ is the Gibbsian
(canonical ensemble) measure on $\mathbb{D}$ possessing the density $\rho (\mathbf{r},%
\mathbf{p},\mathbf{q},\bm{\pi})$ from (\ref{a3}).

Now we will show that $E\varphi (\tilde{X}(t;x_{0}))=E\varphi (X(t;x_{0}))$ for
$\varphi$ which has physical meaning. To this end, introduce
$\bm{\hat \omega}=(\hat{\omega}^{1\,\mathsf{T}},\ldots ,\hat{\omega%
}^{n\,\mathsf{T}})^{\mathsf{T}}\in \mathbb{R}^{4n}$, $\hat{\omega}%
^{j}=(\omega _{0}^{j},\omega _{1}^{j},\omega _{2}^{j},\omega _{3}^{j})^{%
\mathsf{T}}\in \mathbb{R}^{4}$ and $\bm{\omega}=(\omega ^{1\,\mathsf{T}%
},\ldots ,\omega ^{n\,\mathsf{T}})^{\mathsf{T}}\in \mathbb{R}^{3n}$, $\omega
^{j}=(\omega _{1}^{j},\omega _{2}^{j},\omega _{3}^{j})^{\mathsf{T}}\in
\mathbb{R}^{3}$ so that $\hat{\omega}^{j}=\half S^\mathsf{T}(q^{j})\pi ^{j},$ $%
j=1,\ldots ,n.$ We see that $\omega _{1}^{j},$ $\omega _{2}^{j},$ $\omega
_{3}^{j}$ are the (conventional) angular momenta about the axes in the body-centred frame
of molecule $j$, while the components $\omega_{0}^{j}$ are of an axillary
nature. Let $\mathbb{\breve{D}}=\{x=(\mathbf{r}^{\mathsf{T}},\mathbf{p}^{%
\mathsf{T}},\mathbf{q}^{\mathsf{T}},\bm{\pi}^{\mathsf{T}})^{\mathsf{T}}\in
\mathbb{R}^{14n}:\ \ |q^{j}|=1,\ \ j=1,\ldots ,n\}=\{x=(\mathbf{r}^{\mathsf{T%
}},\mathbf{p}^{\mathsf{T}},\mathbf{q}^{\mathsf{T}},\bm{\hat \omega }^{%
\mathsf{T}})^{\mathsf{T}}\in \mathbb{R}^{14n}:\ \ |q^{j}|=1,\ \ j=1,\ldots
,n\}.$ Note that $\mathbb{D}=\{x=(\mathbf{r}^{\mathsf{T}},\mathbf{p}^{%
\mathsf{T}},\mathbf{q}^{\mathsf{T}},\bm{\hat \omega }^{\mathsf{T}})^{\mathsf{%
T}}\in \mathbb{R}^{14n}:\ \ |q^{j}|=1,$ $\omega _{0}^{j}=0,\ \ j=1,\ldots
,n\}.$ Assume that $\varphi (x):~\mathbb{\breve{D}\rightarrow R}$ is a
function with polynomial growth at infinity which does not depend
on the non-physical variable $\bm{\omega}_0=(\omega
_{0}^{1},\ldots ,\omega _{0}^{n})^{\mathsf{T}}.$ It is not difficult to see
that%
\begin{widetext}
\begin{eqnarray}
\varphi (x) &=&\varphi (\mathbf{r}^{\mathsf{T}},\mathbf{p}^{\mathsf{T}},%
\mathbf{q}^{\mathsf{T}},\bm{\pi}^{\mathsf{T}})=\varphi (\mathbf{r}^{\mathsf{T%
}},\mathbf{p}^{\mathsf{T}},\mathbf{q}^{\mathsf{T}},2\hat{\omega}^{1\,\mathsf{T}}
S^{\mathsf{T}}(q^{1}),\ldots ,2\hat{\omega}^{n\,\mathsf{T}}
S^{\mathsf{T}}(q^{n}))  \label{fi1} \\
&=&\varphi (\mathbf{r}^{\mathsf{T}},\mathbf{p}^{\mathsf{T}},\mathbf{q}^{\mathsf{T}},
2(\omega_{0}^{1},\omega_{1}^{1},\omega_{2}^{1},\omega_{3}^{1})S^{\mathsf{T}}(q^{1}),\ldots,
2(\omega_{0}^{n},\omega_{1}^{n},\omega_{2}^{n},\omega_{3}^{n})S^{\mathsf{T}}(q^{n}))  \notag \\
&=&\varphi (\mathbf{r}^{\mathsf{T}},\mathbf{p}^{\mathsf{T}},\mathbf{q}^{\mathsf{T}},
2(0,\omega_{1}^{1},\omega_{2}^{1},\omega_{3}^{1})S^{\mathsf{T}}(q^{1}),\ldots,
2(0,\omega_{1}^{n},\omega_{2}^{n},\omega_{3}^{n})S^{\mathsf{T}}(q^{n}))  \notag \\
&=&\varphi (\mathbf{r}^{\mathsf{T}},\mathbf{p}^{\mathsf{T}},\mathbf{q}^{%
\mathsf{T}},(\pi ^{1}-q^{1\,\mathsf{T}}\pi ^{1}q^{1})^{\mathsf{T}},\ldots
,(\pi ^{n}-q^{n\,\mathsf{T}}\pi ^{n}q^{n})^{\mathsf{T}}).  \notag
\end{eqnarray}%
\end{widetext}
In (\ref{fi1}), we may put $\bm{\omega}_0=0$ since $\varphi (x)$ does not
depend on the non-physical quantity $\bm{\omega}_0$ and any value can be assigned to $\bm{\omega}_0$
without changing the value of $\varphi$.  Further, (\ref{fi1}) implies that
\begin{equation*}
\begin{split}
&E\varphi (\mathbf{R}^{\mathsf{T}},\mathbf{P}^{\mathsf{T}},\mathbf{Q}^{%
\mathsf{T}},\bm{\Pi}^{\mathsf{T}})|_{\operatorname{(\ref{nl1}), (\ref{lt2})}}  \\
&=E\varphi (%
\mathbf{R}^{\mathsf{T}},\mathbf{P}^{\mathsf{T}},\mathbf{Q}^{\mathsf{T}},%
\mathbf{\tilde{\Pi}}^{\mathsf{T}}(t))|_{(\ref{nl1}),(\ref{tilda_pi})},
\end{split}
\end{equation*}%
where expectation on the left is computed with respect to the system (\ref%
{nl1}),~(\ref{lt2}) and on the right with respect to (\ref{nl1}),~(\ref{tilda_pi}).
This explains why (\ref{nl1}),~(\ref{lt2}) can be used for calculating ergodic
limits $\varphi^{erg}$ for functions $\varphi (x)$ of physical interest
(i.e., independent of $q^{j\,\mathsf{T}}\pi ^{j})$ despite (\ref{nl1}),~(\ref%
{lt2}) being not ergodic (cf. (\ref{YY})).

In essence, the evolution of the physical degrees of freedom is unaffected by the non-physical component $\Omega^{j}(t)$, and the physical degrees of freedom sample from the Gibbs measure as desired. Further, no quantity of interest depends on the non-physical component of $\Pi$, and so averages obtained using the old Langevin thermostat
(\ref{nl1}), (\ref{lt2}) are correct.

\section{The stationary Fokker-Planck equation for the gradient thermostat\label{sec:fpe}}
Recall that in (\ref{a10}),~(\ref{ngt1})
\begin{eqnarray*}
f^{j}(\mathbf{r},\mathbf{q})&=& -\nabla _{r^{j}}U(\mathbf{r},\mathbf{q}), \\
\mathbb{F}_{j}(\mathbf{r},\mathbf{q}) &=& F^{j}(\mathbf{r},\mathbf{q})
q^{j\,\mathsf{T}} - q^{j}(F^{j}(\mathbf{r},\mathbf{q}))^{\mathsf{T}}, \\
F^{j}(\mathbf{r},\mathbf{q}) &=&{-\nabla_{q^{j}}U(\mathbf{r},\mathbf{q})+}%
q^{j\,\mathsf{T}}{\nabla _{q^{j}}U(\mathbf{r},\mathbf{q})q^j,}
\end{eqnarray*}%
and hence
\begin{equation*}
\mathbb{F}_{j}(\mathbf{r},\mathbf{q})q^{j}={-\nabla_{q^{j}}
U(\mathbf{r},\mathbf{q})}q^{j\,\mathsf{T}}q^{j}+q^{j\,\mathsf{T}}
\nabla_{q^{j}}U(\mathbf{r},\mathbf{q})q^{j},
\end{equation*}%
where the gradients $\nabla_{r^{j}}$ and ${\nabla_{q^{j}}}$ are in the
Cartesian coordinates in $\mathbb{R}^{3}$ and $\mathbb{R}^{4}$,
respectively.  The stationary Fokker-Planck equation
corresponding to (\ref{a10}),~(\ref{ngt1}) has the form (see, for example,
Refs.~\onlinecite{Has,GC,Soize}):
\begin{equation}
L^{\ast }\rho =0,  \label{fp2}
\end{equation}%
where
\begin{widetext}
\begin{eqnarray*}
L^{\ast }\rho  &=&\sum_{j=1}^{n}\Bigg[ \frac{\upsilon }{m\beta }\sum_{i=1}^{3}%
\frac{\partial ^{2}}{\big( \partial r_{i}^{j}\big) ^{2}}\rho +\frac{%
\Upsilon }{M\beta }\sum_{i,k,l=0}^{3}
\frac{\partial}{\partial q_{i}^{j}}\left\{(S_{l}q^{j})_{i}\frac{%
\partial }{\partial q_{k}^{j}}\left((S_{l}q^{j})_{k}\rho \right)\right\} \\
&& +\frac{\upsilon }{m}\sum_{i=1}^{3}\frac{\partial }{\partial r_{i}^{j}}%
\left( \frac{\partial U(\mathbf{r},\mathbf{q})}{\partial r_{i}^{j}}\rho
\right) +\frac{\Upsilon }{M}\sum_{i=0}^{3}\frac{\partial }{\partial q_{i}^{j}%
}\left\{ \left( \frac{{\partial U(\mathbf{r},\mathbf{q})}}{\partial q_{i}^{j}%
}q^{j\,\mathsf{T}}q^{j}-q^{j\,\mathsf{T}}{\nabla_{q^{j}}U(\mathbf{r},%
\mathbf{q})}q_{i}^{j}\right) \rho \right\} \Bigg] .
\end{eqnarray*}%
\end{widetext}

By direct calculations one can verify that the Gibbsian density $\mathbf{%
\tilde{\rho}(\mathbf{r},\mathbf{q})}$ from (\ref{newden}) satisfies (\ref%
{fp2}). We note that (\ref{fp2}) is written for ${(\mathbf{r},%
\mathbf{q})}\in \mathbb{R}^{7n}$ using the fact that (\ref{a10}), (\ref{ngt1}%
) is defined in $\mathbb{R}^{7n},$ i.e., we do not work here with the
manifold $\mathbb{S}^{3}$ on which $Q^{j}(t)$
from (\ref{ngt1}) naturally live.  Instead we work with $\mathbb{R}^{4}$ in
which $\mathbb{S}^{3}$ is embedded.  As the dynamics of our thermostat and integrator have the property that trajectories initially on the manifold do not leave it, the observation that the Gibbsian density is stationary over $\mathbb{R}^{7n}$ means that it is also an invariant probability measure on the manifold of relevance to our modelling and simulations.

\section{Computing the exponent of a real skew-symmetric matrix of order 4\label{sec:rodr}}
The exponent of a skew-symmetric matrix
\begin{equation*}
A=\left(
\begin{array}{cccc}
0 & u_{6} & u_{5} & u_{3} \\
-u_{6} & 0 & u_{4} & u_{2} \\
-u_{5} & -u_{4} & 0 & u_{1} \\
-u_{3} & -u_{2} & -u_{1} & 0%
\end{array}%
\right) ,
\end{equation*}%
is calculated according to the Rodrigues formula
\begin{equation*}
{\rm e}^{A}=\cos \mu I+\frac{\sin \mu }{\mu }A+(aA+bI)(A^{2}+\mu ^{2}I)\,,
\end{equation*}%
where
\begin{equation*}
a=\frac{\sin \alpha /\alpha -\sin \mu /\mu }{\delta }, \qquad
b=\frac{\cos \alpha -\cos \mu }{\delta }
\end{equation*}%
and
\begin{align*}
\alpha & =\sqrt{{\textstyle\frac{1}{2}}(a_{2}-\delta )}, & \mu & =\sqrt{{%
\textstyle\frac{1}{2}}(a_{2}+\delta )}, \\
\delta & =\sqrt{a_{2}^{2}-4a_{0}}, & & \\
a_{0}& =(u_{1}u_{6}+u_{3}u_{4}-u_{2}u_{5})^{2}, & a_{2}&
=\sum_{i=1}^{6}u_{i}^{2}.
\end{align*}%
This Rodrigues formula was considered in [T. Politi. A formula for the
exponential of a real skew-symmetric matrix of order 4. \textit{BIT Numer.
Math.} 41 (2001), 842--845] but the expression given there contains
misprints, namely signs in the denominator of $a$ and in $c$ are incorrect
in Eq.~(2.5) of Politi's paper.

Note that $a_{0},a_{2}\geq 0$ and $a_{2}^{2}\geq 4a_{0}$, and so $\mu \geq
\alpha $. When $\alpha $ and/or $\mu $, or $\mu -\alpha $ are close to zero,
the evaluation of the above expressions needs to be done with care to avoid
subtractive cancellation or division by zero. In particular, if $\alpha
<10^{-4}$, then replace $\sin \alpha /\alpha $ with $(\alpha
^{2}/20-1)\alpha ^{2}/6+1$. Also replace $\sin \mu /\mu $ with $(\mu
^{2}/20-1)\mu ^{2}/6+1$ when $a_{2}<10^{-8}$.

When $\mu =\alpha $ (i.e. in the limit $\delta \rightarrow 0$), the
eigenvalues of $A$ are degenerate. In this case the expressions for $a$, and
$b$ can be approximated as follows:
\begin{equation*}
a=\frac{\sin \gamma /\gamma -\cos \gamma }{a_{2}}+O(\delta ^{2}),\qquad b=%
\frac{\sin \gamma }{2\gamma }+O(\delta ^{2}),
\end{equation*}%
where $\gamma =\sqrt{a_{2}/2}$. We use this expression when $\delta <10^{-8}$.

If the matrix $A$ is close to zero, i.e., when $a_{2}$ is small, then the
approximations which can be used for $a$ and $b$ are of the form
\begin{equation*}
a={\textstyle\frac{1}{6}}-{\textstyle\frac{1}{120}}a_{2}+O(a_{2}^{2}),\qquad
b={\textstyle\frac{1}{2}}-{\textstyle\frac{1}{24}}a_{2}+O(a_{2}^{2}).
\end{equation*}%
We use these expressions when $a_{2}<10^{-8}$.

\bibliography{langevin}

\begin{thebibliography}{48}%
\makeatletter
\providecommand \@ifxundefined [1]{%
 \@ifx{#1\undefined}
}%
\providecommand \@ifnum [1]{%
 \ifnum #1\expandafter \@firstoftwo
 \else \expandafter \@secondoftwo
 \fi
}%
\providecommand \@ifx [1]{%
 \ifx #1\expandafter \@firstoftwo
 \else \expandafter \@secondoftwo
 \fi
}%
\providecommand \natexlab [1]{#1}%
\providecommand \enquote  [1]{``#1''}%
\providecommand \bibnamefont  [1]{#1}%
\providecommand \bibfnamefont [1]{#1}%
\providecommand \citenamefont [1]{#1}%
\providecommand \href@noop [0]{\@secondoftwo}%
\providecommand \href [0]{\begingroup \@sanitize@url \@href}%
\providecommand \@href[1]{\@@startlink{#1}\@@href}%
\providecommand \@@href[1]{\endgroup#1\@@endlink}%
\providecommand \@sanitize@url [0]{\catcode `\\12\catcode `\$12\catcode
  `\&12\catcode `\#12\catcode `\^12\catcode `\_12\catcode `\%12\relax}%
\providecommand \@@startlink[1]{}%
\providecommand \@@endlink[0]{}%
\providecommand \url  [0]{\begingroup\@sanitize@url \@url }%
\providecommand \@url [1]{\endgroup\@href {#1}{\urlprefix }}%
\providecommand \urlprefix  [0]{URL }%
\providecommand \Eprint [0]{\href }%
\providecommand \doibase [0]{http://dx.doi.org/}%
\providecommand \selectlanguage [0]{\@gobble}%
\providecommand \bibinfo  [0]{\@secondoftwo}%
\providecommand \bibfield  [0]{\@secondoftwo}%
\providecommand \translation [1]{[#1]}%
\providecommand \BibitemOpen [0]{}%
\providecommand \bibitemStop [0]{}%
\providecommand \bibitemNoStop [0]{.\EOS\space}%
\providecommand \EOS [0]{\spacefactor3000\relax}%
\providecommand \BibitemShut  [1]{\csname bibitem#1\endcsname}%
\let\auto@bib@innerbib\@empty
\bibitem [{\citenamefont {Allen}\ and\ \citenamefont
  {Tildesley}(1987)}]{Allen}%
  \BibitemOpen
  \bibfield  {author} {\bibinfo {author} {\bibfnamefont {M.}~\bibnamefont
  {Allen}}\ and\ \bibinfo {author} {\bibfnamefont {D.}~\bibnamefont
  {Tildesley}},\ }\href@noop {} {\emph {\bibinfo {title} {Computer Simulation
  of Liquids}}}\ (\bibinfo  {publisher} {Oxford University Press},\ \bibinfo
  {address} {Oxford},\ \bibinfo {year} {1987})\BibitemShut {NoStop}%
\bibitem [{\citenamefont {Leimkuhler}\ and\ \citenamefont {Reich}(2005)}]{Ben}%
  \BibitemOpen
  \bibfield  {author} {\bibinfo {author} {\bibfnamefont {B.}~\bibnamefont
  {Leimkuhler}}\ and\ \bibinfo {author} {\bibfnamefont {S.}~\bibnamefont
  {Reich}},\ }\href@noop {} {\emph {\bibinfo {title} {Simulating {H}amiltonian
  Dynamics}}}\ (\bibinfo  {publisher} {Cambridge University Press},\ \bibinfo
  {address} {Cambridge},\ \bibinfo {year} {2005})\BibitemShut {NoStop}%
\bibitem [{\citenamefont {Schlick}(2010)}]{Schlick}%
  \BibitemOpen
  \bibfield  {author} {\bibinfo {author} {\bibfnamefont {T.}~\bibnamefont
  {Schlick}},\ }\href@noop {} {\emph {\bibinfo {title} {Molecular Modeling and
  Simulation: An Interdisciplinary Guide}}},\ \bibinfo {edition} {2nd}\ ed.\
  (\bibinfo  {publisher} {Springer},\ \bibinfo {address} {New York},\ \bibinfo
  {year} {2010})\BibitemShut {NoStop}%
\bibitem [{\citenamefont {Snook}(2006)}]{Snook}%
  \BibitemOpen
  \bibfield  {author} {\bibinfo {author} {\bibfnamefont {I.}~\bibnamefont
  {Snook}},\ }\href@noop {} {\emph {\bibinfo {title} {The {L}angevin and
  Generalised {L}angevin Approach to the Dynamics of Atomic, Polymeric and
  Colloidal Systems}}}\ (\bibinfo  {publisher} {Elsevier},\ \bibinfo {year}
  {2006})\BibitemShut {NoStop}%
\bibitem [{\citenamefont {Davidchack}, \citenamefont {Handel},\ and\
  \citenamefont {Tretyakov}(2009)}]{DHT09}%
  \BibitemOpen
  \bibfield  {author} {\bibinfo {author} {\bibfnamefont {R.~L.}\ \bibnamefont
  {Davidchack}}, \bibinfo {author} {\bibfnamefont {R.}~\bibnamefont {Handel}},
  \ and\ \bibinfo {author} {\bibfnamefont {M.~V.}\ \bibnamefont {Tretyakov}},\
  }\href@noop {} {\bibfield  {journal} {\bibinfo  {journal} {J. Chem. Phys.}\
  }\textbf {\bibinfo {volume} {130}},\ \bibinfo {pages} {234101} (\bibinfo
  {year} {2009})}\BibitemShut {NoStop}%
\bibitem [{\citenamefont {Milstein}\ and\ \citenamefont
  {Tretyakov}(2007)}]{MT7}%
  \BibitemOpen
  \bibfield  {author} {\bibinfo {author} {\bibfnamefont {G.~N.}\ \bibnamefont
  {Milstein}}\ and\ \bibinfo {author} {\bibfnamefont {M.~V.}\ \bibnamefont
  {Tretyakov}},\ }\href@noop {} {\bibfield  {journal} {\bibinfo  {journal}
  {Physica D}\ }\textbf {\bibinfo {volume} {229}},\ \bibinfo {pages} {81}
  (\bibinfo {year} {2007})}\BibitemShut {NoStop}%
\bibitem [{\citenamefont {Bulgac}\ and\ \citenamefont
  {Kusnezov}(1990)}]{Bulgac90}%
  \BibitemOpen
  \bibfield  {author} {\bibinfo {author} {\bibfnamefont {A.}~\bibnamefont
  {Bulgac}}\ and\ \bibinfo {author} {\bibfnamefont {D.}~\bibnamefont
  {Kusnezov}},\ }\href@noop {} {\bibfield  {journal} {\bibinfo  {journal}
  {Phys. Rev. A}\ }\textbf {\bibinfo {volume} {42}},\ \bibinfo {pages} {5045}
  (\bibinfo {year} {1990})}\BibitemShut {NoStop}%
\bibitem [{\citenamefont {Samoletov}, \citenamefont {Dettmann},\ and\
  \citenamefont {Chaplain}(2007)}]{Samoletov07}%
  \BibitemOpen
  \bibfield  {author} {\bibinfo {author} {\bibfnamefont {A.~A.}\ \bibnamefont
  {Samoletov}}, \bibinfo {author} {\bibfnamefont {C.~P.}\ \bibnamefont
  {Dettmann}}, \ and\ \bibinfo {author} {\bibfnamefont {M.~A.~J.}\ \bibnamefont
  {Chaplain}},\ }\href@noop {} {\bibfield  {journal} {\bibinfo  {journal} {J.
  Stat. Phys.}\ }\textbf {\bibinfo {volume} {128}},\ \bibinfo {pages} {1321}
  (\bibinfo {year} {2007})}\BibitemShut {NoStop}%
\bibitem [{\citenamefont {Leimkuhler}(2010)}]{BL10}%
  \BibitemOpen
  \bibfield  {author} {\bibinfo {author} {\bibfnamefont {B.}~\bibnamefont
  {Leimkuhler}},\ }\href@noop {} {\bibfield  {journal} {\bibinfo  {journal}
  {Phys. Rev. E}\ }\textbf {\bibinfo {volume} {81}},\ \bibinfo {pages} {026703}
  (\bibinfo {year} {2010})}\BibitemShut {NoStop}%
\bibitem [{\citenamefont {{Miller III}}\ \emph {et~al.}(2002)\citenamefont
  {{Miller III}}, \citenamefont {Eleftheriou}, \citenamefont {Pattnaik},
  \citenamefont {Ndirango}, \citenamefont {Newns},\ and\ \citenamefont
  {Martyna}}]{qua02}%
  \BibitemOpen
  \bibfield  {author} {\bibinfo {author} {\bibfnamefont {T.~F.}\ \bibnamefont
  {{Miller III}}}, \bibinfo {author} {\bibfnamefont {M.}~\bibnamefont
  {Eleftheriou}}, \bibinfo {author} {\bibfnamefont {P.}~\bibnamefont
  {Pattnaik}}, \bibinfo {author} {\bibfnamefont {A.}~\bibnamefont {Ndirango}},
  \bibinfo {author} {\bibfnamefont {D.}~\bibnamefont {Newns}}, \ and\ \bibinfo
  {author} {\bibfnamefont {G.~J.}\ \bibnamefont {Martyna}},\ }\href@noop {}
  {\bibfield  {journal} {\bibinfo  {journal} {J. Chem. Phys.}\ }\textbf
  {\bibinfo {volume} {116}},\ \bibinfo {pages} {8649} (\bibinfo {year}
  {2002})}\BibitemShut {NoStop}%
\bibitem [{\citenamefont {Evans}(1977)}]{Evans77a}%
  \BibitemOpen
  \bibfield  {author} {\bibinfo {author} {\bibfnamefont {D.~J.}\ \bibnamefont
  {Evans}},\ }\href@noop {} {\bibfield  {journal} {\bibinfo  {journal} {Mol.
  Phys.}\ }\textbf {\bibinfo {volume} {34}},\ \bibinfo {pages} {317} (\bibinfo
  {year} {1977})}\BibitemShut {NoStop}%
\bibitem [{\citenamefont {Evans}\ and\ \citenamefont {Murad}(1977)}]{Evans77b}%
  \BibitemOpen
  \bibfield  {author} {\bibinfo {author} {\bibfnamefont {D.~J.}\ \bibnamefont
  {Evans}}\ and\ \bibinfo {author} {\bibfnamefont {S.}~\bibnamefont {Murad}},\
  }\href@noop {} {\bibfield  {journal} {\bibinfo  {journal} {Mol. Phys.}\
  }\textbf {\bibinfo {volume} {34}},\ \bibinfo {pages} {327} (\bibinfo {year}
  {1977})}\BibitemShut {NoStop}%
\bibitem [{\citenamefont {Dullweber}, \citenamefont {Leimkuhler},\ and\
  \citenamefont {McLachlan}(1997)}]{DLM97}%
  \BibitemOpen
  \bibfield  {author} {\bibinfo {author} {\bibfnamefont {A.}~\bibnamefont
  {Dullweber}}, \bibinfo {author} {\bibfnamefont {B.}~\bibnamefont
  {Leimkuhler}}, \ and\ \bibinfo {author} {\bibfnamefont {R.}~\bibnamefont
  {McLachlan}},\ }\href@noop {} {\bibfield  {journal} {\bibinfo  {journal} {J.
  Chem. Phys.}\ }\textbf {\bibinfo {volume} {107}},\ \bibinfo {pages} {5840}
  (\bibinfo {year} {1997})}\BibitemShut {NoStop}%
\bibitem [{\citenamefont {Omelyan}(2007)}]{Ome07}%
  \BibitemOpen
  \bibfield  {author} {\bibinfo {author} {\bibfnamefont {I.~P.}\ \bibnamefont
  {Omelyan}},\ }\href@noop {} {\bibfield  {journal} {\bibinfo  {journal} {J.
  Chem. Phys.}\ }\textbf {\bibinfo {volume} {127}},\ \bibinfo {pages} {044102}
  (\bibinfo {year} {2007})}\BibitemShut {NoStop}%
\bibitem [{\citenamefont {van {Z}on}, \citenamefont {Omelyan},\ and\
  \citenamefont {Schofield}(2008)}]{vanZon08jcp}%
  \BibitemOpen
  \bibfield  {author} {\bibinfo {author} {\bibfnamefont {R.}~\bibnamefont {van
  {Z}on}}, \bibinfo {author} {\bibfnamefont {I.~P.}\ \bibnamefont {Omelyan}}, \
  and\ \bibinfo {author} {\bibfnamefont {J.}~\bibnamefont {Schofield}},\
  }\href@noop {} {\bibfield  {journal} {\bibinfo  {journal} {J. Chem. Phys.}\
  }\textbf {\bibinfo {volume} {128}},\ \bibinfo {pages} {136102} (\bibinfo
  {year} {2008})}\BibitemShut {NoStop}%
\bibitem [{\citenamefont {Omelyan}(2008)}]{Omelyan08pre}%
  \BibitemOpen
  \bibfield  {author} {\bibinfo {author} {\bibfnamefont {I.~P.}\ \bibnamefont
  {Omelyan}},\ }\href@noop {} {\bibfield  {journal} {\bibinfo  {journal} {Phys.
  Rev. E}\ }\textbf {\bibinfo {volume} {78}},\ \bibinfo {pages} {026702}
  (\bibinfo {year} {2008})}\BibitemShut {NoStop}%
\bibitem [{\citenamefont {Milstein}\ and\ \citenamefont
  {Tretyakov}(2004)}]{MT1}%
  \BibitemOpen
  \bibfield  {author} {\bibinfo {author} {\bibfnamefont {G.~N.}\ \bibnamefont
  {Milstein}}\ and\ \bibinfo {author} {\bibfnamefont {M.~V.}\ \bibnamefont
  {Tretyakov}},\ }\href@noop {} {\emph {\bibinfo {title} {Stochastic Numerics
  for Mathematical Physics}}}\ (\bibinfo  {publisher} {Springer},\ \bibinfo
  {address} {Berlin},\ \bibinfo {year} {2004})\BibitemShut {NoStop}%
\bibitem [{\citenamefont {Ouldridge}\ \emph
  {et~al.}(2013{\natexlab{a}})\citenamefont {Ouldridge}, \citenamefont {Hoare},
  \citenamefont {Louis}, \citenamefont {Doye}, \citenamefont {Bath},\ and\
  \citenamefont {Turberfield}}]{Ouldridge_walker_2013}%
  \BibitemOpen
  \bibfield  {author} {\bibinfo {author} {\bibfnamefont {T.~E.}\ \bibnamefont
  {Ouldridge}}, \bibinfo {author} {\bibfnamefont {R.~L.}\ \bibnamefont
  {Hoare}}, \bibinfo {author} {\bibfnamefont {A.~A.}\ \bibnamefont {Louis}},
  \bibinfo {author} {\bibfnamefont {J.~P.~K.}\ \bibnamefont {Doye}}, \bibinfo
  {author} {\bibfnamefont {J.}~\bibnamefont {Bath}}, \ and\ \bibinfo {author}
  {\bibfnamefont {A.~J.}\ \bibnamefont {Turberfield}},\ }\href@noop {}
  {\bibfield  {journal} {\bibinfo  {journal} {ACS Nano}\ }\textbf {\bibinfo
  {volume} {7}},\ \bibinfo {pages} {2479} (\bibinfo {year}
  {2013}{\natexlab{a}})}\BibitemShut {NoStop}%
\bibitem [{\citenamefont {Ouldridge}\ \emph
  {et~al.}(2013{\natexlab{b}})\citenamefont {Ouldridge}, \citenamefont
  {\v{S}ulc}, \citenamefont {Romano}, \citenamefont {Doye},\ and\ \citenamefont
  {Louis}}]{Ouldridge_binding_2013}%
  \BibitemOpen
  \bibfield  {author} {\bibinfo {author} {\bibfnamefont {T.~E.}\ \bibnamefont
  {Ouldridge}}, \bibinfo {author} {\bibfnamefont {P.}~\bibnamefont {\v{S}ulc}},
  \bibinfo {author} {\bibfnamefont {F.}~\bibnamefont {Romano}}, \bibinfo
  {author} {\bibfnamefont {J.~P.~K.}\ \bibnamefont {Doye}}, \ and\ \bibinfo
  {author} {\bibfnamefont {A.~A.}\ \bibnamefont {Louis}},\ }\href@noop {}
  {\bibfield  {journal} {\bibinfo  {journal} {Nucl. Acids Res.}\ }\textbf
  {\bibinfo {volume} {41}},\ \bibinfo {pages} {8886} (\bibinfo {year}
  {2013}{\natexlab{b}})}\BibitemShut {NoStop}%
\bibitem [{\citenamefont {Srinivas}\ \emph {et~al.}(2013)\citenamefont
  {Srinivas}, \citenamefont {Ouldridge}, \citenamefont {\v{S}ulc},
  \citenamefont {Schaeffer}, \citenamefont {Yurke}, \citenamefont {Louis},
  \citenamefont {Doye},\ and\ \citenamefont {Winfree}}]{Srinivas2013}%
  \BibitemOpen
  \bibfield  {author} {\bibinfo {author} {\bibfnamefont {N.}~\bibnamefont
  {Srinivas}}, \bibinfo {author} {\bibfnamefont {T.~E.}\ \bibnamefont
  {Ouldridge}}, \bibinfo {author} {\bibfnamefont {P.}~\bibnamefont {\v{S}ulc}},
  \bibinfo {author} {\bibfnamefont {J.}~\bibnamefont {Schaeffer}}, \bibinfo
  {author} {\bibfnamefont {B.}~\bibnamefont {Yurke}}, \bibinfo {author}
  {\bibfnamefont {A.~A.}\ \bibnamefont {Louis}}, \bibinfo {author}
  {\bibfnamefont {J.~P.~K.}\ \bibnamefont {Doye}}, \ and\ \bibinfo {author}
  {\bibfnamefont {E.}~\bibnamefont {Winfree}},\ }\href@noop {} {\bibfield
  {journal} {\bibinfo  {journal} {Nucl. Acids Res.}\ }\textbf {\bibinfo
  {volume} {41}},\ \bibinfo {pages} {10641} (\bibinfo {year}
  {2013})}\BibitemShut {NoStop}%
\bibitem [{\citenamefont {Machinek}\ \emph {et~al.}(2014)\citenamefont
  {Machinek}, \citenamefont {Ouldridge}, \citenamefont {Haley}, \citenamefont
  {Bath},\ and\ \citenamefont {Turberfield}}]{Machinek2014}%
  \BibitemOpen
  \bibfield  {author} {\bibinfo {author} {\bibfnamefont {R.~R.~F.}\
  \bibnamefont {Machinek}}, \bibinfo {author} {\bibfnamefont {T.~E.}\
  \bibnamefont {Ouldridge}}, \bibinfo {author} {\bibfnamefont {N.~E.~C.}\
  \bibnamefont {Haley}}, \bibinfo {author} {\bibfnamefont {J.}~\bibnamefont
  {Bath}}, \ and\ \bibinfo {author} {\bibfnamefont {A.~J.}\ \bibnamefont
  {Turberfield}},\ }\href@noop {} {\bibfield  {journal} {\bibinfo  {journal}
  {Nat. Commun.}\ }\textbf {\bibinfo {volume} {5}},\ \bibinfo {pages} {5324}
  (\bibinfo {year} {2014})}\BibitemShut {NoStop}%
\bibitem [{\citenamefont {Vanden-Eijnden}\ and\ \citenamefont
  {Ciccotti}(2006)}]{Eric06}%
  \BibitemOpen
  \bibfield  {author} {\bibinfo {author} {\bibfnamefont {E.}~\bibnamefont
  {Vanden-Eijnden}}\ and\ \bibinfo {author} {\bibfnamefont {G.}~\bibnamefont
  {Ciccotti}},\ }\href@noop {} {\bibfield  {journal} {\bibinfo  {journal}
  {Chem. Phys. Lett.}\ }\textbf {\bibinfo {volume} {429}},\ \bibinfo {pages}
  {310�} (\bibinfo {year} {2006})}\BibitemShut {NoStop}%
\bibitem [{\citenamefont {Sun}, \citenamefont {Lin},\ and\ \citenamefont
  {Gezelter}(2008)}]{Sun08}%
  \BibitemOpen
  \bibfield  {author} {\bibinfo {author} {\bibfnamefont {X.}~\bibnamefont
  {Sun}}, \bibinfo {author} {\bibfnamefont {T.}~\bibnamefont {Lin}}, \ and\
  \bibinfo {author} {\bibfnamefont {J.~D.}\ \bibnamefont {Gezelter}},\
  }\href@noop {} {\bibfield  {journal} {\bibinfo  {journal} {J. Chem. Phys.}\
  }\textbf {\bibinfo {volume} {128}},\ \bibinfo {pages} {234107} (\bibinfo
  {year} {2008})}\BibitemShut {NoStop}%
\bibitem [{\citenamefont {Altmann}(1986)}]{quabook}%
  \BibitemOpen
  \bibfield  {author} {\bibinfo {author} {\bibfnamefont {S.~L.}\ \bibnamefont
  {Altmann}},\ }\href@noop {} {\emph {\bibinfo {title} {Rotations, Quaternions,
  and Double Groups}}}\ (\bibinfo  {publisher} {Dover Publications},\ \bibinfo
  {address} {New York},\ \bibinfo {year} {1986})\BibitemShut {NoStop}%
\bibitem [{\citenamefont {Heard}(2006)}]{quabook2}%
  \BibitemOpen
  \bibfield  {author} {\bibinfo {author} {\bibfnamefont {W.~B.}\ \bibnamefont
  {Heard}},\ }\href@noop {} {\emph {\bibinfo {title} {Rigid Body Mechanics}}}\
  (\bibinfo  {publisher} {Wiley},\ \bibinfo {address} {Weinheim},\ \bibinfo
  {year} {2006})\BibitemShut {NoStop}%
\bibitem [{Note1()}]{Note1}%
  \BibitemOpen
  \bibinfo {note} {Note that in Eq.~(3) of Ref.~\protect \rev@citealpnum
  {DHT09} the notation $\nabla _{q^{j}}$ for the directional derivative was
  used in $\nabla _{q^{j}}U$ while $\nabla _{q^{j}}V$ meant the conventional
  gradient.}\BibitemShut {Stop}%
\bibitem [{\citenamefont {Hasminskii}(1980)}]{Has}%
  \BibitemOpen
  \bibfield  {author} {\bibinfo {author} {\bibfnamefont {R.~Z.}\ \bibnamefont
  {Hasminskii}},\ }\href@noop {} {\emph {\bibinfo {title} {Stochastic Stability
  of Differential Equations}}}\ (\bibinfo  {publisher} {Sijthoff \&
  Noordhoff},\ \bibinfo {year} {1980})\BibitemShut {NoStop}%
\bibitem [{\citenamefont {Soize}(1994)}]{Soize}%
  \BibitemOpen
  \bibfield  {author} {\bibinfo {author} {\bibfnamefont {C.}~\bibnamefont
  {Soize}},\ }\href@noop {} {\emph {\bibinfo {title} {The {F}okker-{P}lanck
  Equation for Stochastic Dynamical Systems and Its Explicit Steady State
  Solutions}}}\ (\bibinfo  {publisher} {World Scientific},\ \bibinfo {address}
  {Singapore},\ \bibinfo {year} {1994})\BibitemShut {NoStop}%
\bibitem [{\citenamefont {Rossky}, \citenamefont {Doll},\ and\ \citenamefont
  {Friedman}(1978)}]{Doll87}%
  \BibitemOpen
  \bibfield  {author} {\bibinfo {author} {\bibfnamefont {P.~J.}\ \bibnamefont
  {Rossky}}, \bibinfo {author} {\bibfnamefont {J.~D.}\ \bibnamefont {Doll}}, \
  and\ \bibinfo {author} {\bibfnamefont {H.~L.}\ \bibnamefont {Friedman}},\
  }\href@noop {} {\bibfield  {journal} {\bibinfo  {journal} {J. Chem. Phys.}\
  }\textbf {\bibinfo {volume} {69}},\ \bibinfo {pages} {4628} (\bibinfo {year}
  {1978})}\BibitemShut {NoStop}%
\bibitem [{\citenamefont {Leimkuhler}\ and\ \citenamefont
  {Matthews}(2013)}]{BenM13}%
  \BibitemOpen
  \bibfield  {author} {\bibinfo {author} {\bibfnamefont {B.}~\bibnamefont
  {Leimkuhler}}\ and\ \bibinfo {author} {\bibfnamefont {C.}~\bibnamefont
  {Matthews}},\ }\href@noop {} {\bibfield  {journal} {\bibinfo  {journal}
  {Appl. Math. Res. Express}\ }\textbf {\bibinfo {volume} {2013}},\ \bibinfo
  {pages} {34} (\bibinfo {year} {2013})}\BibitemShut {NoStop}%
\bibitem [{\citenamefont {Hairer}, \citenamefont {Lubich},\ and\ \citenamefont
  {Wanner}(2006)}]{HLW02}%
  \BibitemOpen
  \bibfield  {author} {\bibinfo {author} {\bibfnamefont {E.}~\bibnamefont
  {Hairer}}, \bibinfo {author} {\bibfnamefont {C.}~\bibnamefont {Lubich}}, \
  and\ \bibinfo {author} {\bibfnamefont {G.}~\bibnamefont {Wanner}},\
  }\href@noop {} {\emph {\bibinfo {title} {Geometric Numerical Integration}}}\
  (\bibinfo  {publisher} {Springer Ser. Comput. Math. 31, Springer-Verlag},\
  \bibinfo {address} {Berlin},\ \bibinfo {year} {2006})\BibitemShut {NoStop}%
\bibitem [{\citenamefont {Milstein}, \citenamefont {Repin},\ and\ \citenamefont
  {Tretyakov}(2002)}]{MilRT02}%
  \BibitemOpen
  \bibfield  {author} {\bibinfo {author} {\bibfnamefont {G.~N.}\ \bibnamefont
  {Milstein}}, \bibinfo {author} {\bibfnamefont {Y.~M.}\ \bibnamefont {Repin}},
  \ and\ \bibinfo {author} {\bibfnamefont {M.~V.}\ \bibnamefont {Tretyakov}},\
  }\href@noop {} {\bibfield  {journal} {\bibinfo  {journal} {SIAM J. Numer.
  Anal.}\ }\textbf {\bibinfo {volume} {40}},\ \bibinfo {pages} {1583} (\bibinfo
  {year} {2002})}\BibitemShut {NoStop}%
\bibitem [{\citenamefont {Milstein}\ and\ \citenamefont
  {Tretyakov}(2003)}]{MT03}%
  \BibitemOpen
  \bibfield  {author} {\bibinfo {author} {\bibfnamefont {G.~N.}\ \bibnamefont
  {Milstein}}\ and\ \bibinfo {author} {\bibfnamefont {M.~V.}\ \bibnamefont
  {Tretyakov}},\ }\href@noop {} {\bibfield  {journal} {\bibinfo  {journal} {IMA
  J. Numer. Anal.}\ }\textbf {\bibinfo {volume} {23}},\ \bibinfo {pages} {593}
  (\bibinfo {year} {2003})}\BibitemShut {NoStop}%
\bibitem [{\citenamefont {Gr{{\o}}nbech-Jensen}\ and\ \citenamefont
  {Farago}(2013)}]{Referee2}%
  \BibitemOpen
  \bibfield  {author} {\bibinfo {author} {\bibfnamefont {N.}~\bibnamefont
  {Gr{{\o}}nbech-Jensen}}\ and\ \bibinfo {author} {\bibfnamefont
  {O.}~\bibnamefont {Farago}},\ }\href@noop {} {\bibfield  {journal} {\bibinfo
  {journal} {Mol. Phys.}\ }\textbf {\bibinfo {volume} {111}},\ \bibinfo {pages}
  {983} (\bibinfo {year} {2013})}\BibitemShut {NoStop}%
\bibitem [{\citenamefont {Skeel}(1999)}]{Skeel99}%
  \BibitemOpen
  \bibfield  {author} {\bibinfo {author} {\bibfnamefont {R.~D.}\ \bibnamefont
  {Skeel}},\ }in\ \href@noop {} {\emph {\bibinfo {booktitle} {Graduate
  Student's Guide to Numerical Analysis'98}}},\ \bibinfo {editor} {edited by\
  \bibinfo {editor} {\bibfnamefont {M.}~\bibnamefont {Ainsworth}}, \bibinfo
  {editor} {\bibfnamefont {J.}~\bibnamefont {Levesley}}, \ and\ \bibinfo
  {editor} {\bibfnamefont {M.}~\bibnamefont {Marletta}}}\ (\bibinfo
  {publisher} {Springer},\ \bibinfo {year} {1999})\ pp.\ \bibinfo {pages}
  {118--176}\BibitemShut {NoStop}%
\bibitem [{\citenamefont {Castell}\ and\ \citenamefont
  {Gaines}(1995)}]{CasGa95}%
  \BibitemOpen
  \bibfield  {author} {\bibinfo {author} {\bibfnamefont {F.}~\bibnamefont
  {Castell}}\ and\ \bibinfo {author} {\bibfnamefont {J.}~\bibnamefont
  {Gaines}},\ }\href@noop {} {\bibfield  {journal} {\bibinfo  {journal} {Math.
  Comp. Simulation}\ }\textbf {\bibinfo {volume} {38}},\ \bibinfo {pages} {13}
  (\bibinfo {year} {1995})}\BibitemShut {NoStop}%
\bibitem [{\citenamefont {Ikeda}\ and\ \citenamefont {Watanabe}(1981)}]{IkWa}%
  \BibitemOpen
  \bibfield  {author} {\bibinfo {author} {\bibfnamefont {N.}~\bibnamefont
  {Ikeda}}\ and\ \bibinfo {author} {\bibfnamefont {S.}~\bibnamefont
  {Watanabe}},\ }\href@noop {} {\emph {\bibinfo {title} {Stochastic
  Differential Equations and Diffusion Processes}}}\ (\bibinfo  {publisher}
  {North-Holland},\ \bibinfo {address} {Amsterdam},\ \bibinfo {year}
  {1981})\BibitemShut {NoStop}%
\bibitem [{\citenamefont {Elworthy}(1982)}]{Elw}%
  \BibitemOpen
  \bibfield  {author} {\bibinfo {author} {\bibfnamefont {K.~D.}\ \bibnamefont
  {Elworthy}},\ }\href@noop {} {\emph {\bibinfo {title} {Stochastic
  differential equations on manifolds}}},\ London Mathematical Society Lecture
  Note Series 70\ (\bibinfo  {publisher} {Cambridge University Press},\
  \bibinfo {address} {Cambridge},\ \bibinfo {year} {1982})\BibitemShut
  {NoStop}%
\bibitem [{\citenamefont {Rogers}\ and\ \citenamefont
  {Williams}(2000)}]{WilRog}%
  \BibitemOpen
  \bibfield  {author} {\bibinfo {author} {\bibfnamefont {L.~C.~G.}\
  \bibnamefont {Rogers}}\ and\ \bibinfo {author} {\bibfnamefont
  {D.}~\bibnamefont {Williams}},\ }\href@noop {} {\emph {\bibinfo {title}
  {Diffusions, {M}arkov processes, and martingales. {V}ol. 1}}}\ (\bibinfo
  {publisher} {Cambridge University Press},\ \bibinfo {address} {Cambridge},\
  \bibinfo {year} {2000})\BibitemShut {NoStop}%
\bibitem [{\citenamefont {Chirikjian}(2009)}]{GC}%
  \BibitemOpen
  \bibfield  {author} {\bibinfo {author} {\bibfnamefont {G.~S.}\ \bibnamefont
  {Chirikjian}},\ }\href@noop {} {\emph {\bibinfo {title} {Stochastic Models,
  Information Theory, and {L}ie Groups. Volume 1}}},\ Applied and numerical
  harmonic analysis\ (\bibinfo  {publisher} {Birkh{\"{a}}user},\ \bibinfo
  {address} {Basel},\ \bibinfo {year} {2009})\BibitemShut {NoStop}%
\bibitem [{\citenamefont {Davidchack}(2010)}]{RLD10}%
  \BibitemOpen
  \bibfield  {author} {\bibinfo {author} {\bibfnamefont {R.~L.}\ \bibnamefont
  {Davidchack}},\ }\href@noop {} {\bibfield  {journal} {\bibinfo  {journal} {J.
  Comput. Phys.}\ }\textbf {\bibinfo {volume} {229}},\ \bibinfo {pages} {9323}
  (\bibinfo {year} {2010})}\BibitemShut {NoStop}%
\bibitem [{\citenamefont {Chialvo}\ \emph {et~al.}(2001)\citenamefont
  {Chialvo}, \citenamefont {Simonson}, \citenamefont {Cummings},\ and\
  \citenamefont {Kusalik}}]{Chialvo01}%
  \BibitemOpen
  \bibfield  {author} {\bibinfo {author} {\bibfnamefont {A.~A.}\ \bibnamefont
  {Chialvo}}, \bibinfo {author} {\bibfnamefont {J.~M.}\ \bibnamefont
  {Simonson}}, \bibinfo {author} {\bibfnamefont {P.~T.}\ \bibnamefont
  {Cummings}}, \ and\ \bibinfo {author} {\bibfnamefont {P.~G.}\ \bibnamefont
  {Kusalik}},\ }\href@noop {} {\bibfield  {journal} {\bibinfo  {journal} {J.
  Chem. Phys.}\ }\textbf {\bibinfo {volume} {114}},\ \bibinfo {pages} {6514}
  (\bibinfo {year} {2001})}\BibitemShut {NoStop}%
\bibitem [{\citenamefont {Talay}\ and\ \citenamefont {Tubaro}(1990)}]{TAT90}%
  \BibitemOpen
  \bibfield  {author} {\bibinfo {author} {\bibfnamefont {D.}~\bibnamefont
  {Talay}}\ and\ \bibinfo {author} {\bibfnamefont {L.}~\bibnamefont {Tubaro}},\
  }\href@noop {} {\bibfield  {journal} {\bibinfo  {journal} {Stoch. Anal.
  Appl.}\ }\textbf {\bibinfo {volume} {8}},\ \bibinfo {pages} {483} (\bibinfo
  {year} {1990})}\BibitemShut {NoStop}%
\bibitem [{\citenamefont {Frenkel}\ and\ \citenamefont
  {Smit}(2002)}]{FrenkelBook}%
  \BibitemOpen
  \bibfield  {author} {\bibinfo {author} {\bibfnamefont {D.}~\bibnamefont
  {Frenkel}}\ and\ \bibinfo {author} {\bibfnamefont {B.}~\bibnamefont {Smit}},\
  }\href@noop {} {\emph {\bibinfo {title} {Understanding Molecular
  Simulation}}},\ \bibinfo {edition} {2nd}\ ed.\ (\bibinfo  {publisher}
  {Academic Press},\ \bibinfo {address} {New York},\ \bibinfo {year}
  {2002})\BibitemShut {NoStop}%
\bibitem [{\citenamefont {Mattingly}, \citenamefont {Stuart},\ and\
  \citenamefont {Tretyakov}(2010)}]{MST10}%
  \BibitemOpen
  \bibfield  {author} {\bibinfo {author} {\bibfnamefont {J.~C.}\ \bibnamefont
  {Mattingly}}, \bibinfo {author} {\bibfnamefont {A.~M.}\ \bibnamefont
  {Stuart}}, \ and\ \bibinfo {author} {\bibfnamefont {M.~V.}\ \bibnamefont
  {Tretyakov}},\ }\href@noop {} {\bibfield  {journal} {\bibinfo  {journal}
  {SIAM J. Numer. Anal.}\ }\textbf {\bibinfo {volume} {48}},\ \bibinfo {pages}
  {552} (\bibinfo {year} {2010})}\BibitemShut {NoStop}%
\bibitem [{\citenamefont {Mentink}\ \emph {et~al.}(2010)\citenamefont
  {Mentink}, \citenamefont {Tretyakov}, \citenamefont {Fasolino}, \citenamefont
  {Katsnelson},\ and\ \citenamefont {Rasing}}]{Johan10}%
  \BibitemOpen
  \bibfield  {author} {\bibinfo {author} {\bibfnamefont {J.~H.}\ \bibnamefont
  {Mentink}}, \bibinfo {author} {\bibfnamefont {M.~V.}\ \bibnamefont
  {Tretyakov}}, \bibinfo {author} {\bibfnamefont {A.}~\bibnamefont {Fasolino}},
  \bibinfo {author} {\bibfnamefont {M.~I.}\ \bibnamefont {Katsnelson}}, \ and\
  \bibinfo {author} {\bibfnamefont {T.}~\bibnamefont {Rasing}},\ }\href@noop {}
  {\bibfield  {journal} {\bibinfo  {journal} {J. Phys.: Condens. Matter}\
  }\textbf {\bibinfo {volume} {22}},\ \bibinfo {pages} {176001} (\bibinfo
  {year} {2010})}\BibitemShut {NoStop}%
\bibitem [{\citenamefont {Leimkuhler}, \citenamefont {Matthews},\ and\
  \citenamefont {Tretyakov}(2014)}]{LMT14}%
  \BibitemOpen
  \bibfield  {author} {\bibinfo {author} {\bibfnamefont {B.}~\bibnamefont
  {Leimkuhler}}, \bibinfo {author} {\bibfnamefont {C.}~\bibnamefont
  {Matthews}}, \ and\ \bibinfo {author} {\bibfnamefont {M.~V.}\ \bibnamefont
  {Tretyakov}},\ }\href@noop {} {\bibfield  {journal} {\bibinfo  {journal}
  {Proc. R. Soc. A}\ }\textbf {\bibinfo {volume} {470}},\ \bibinfo {pages}
  {20140120} (\bibinfo {year} {2014})}\BibitemShut {NoStop}%
\bibitem [{\citenamefont {Gr{{\o}}nbech-Jensen}\ and\ \citenamefont
  {Farago}(2014)}]{Referee2b}%
  \BibitemOpen
  \bibfield  {author} {\bibinfo {author} {\bibfnamefont {N.}~\bibnamefont
  {Gr{{\o}}nbech-Jensen}}\ and\ \bibinfo {author} {\bibfnamefont
  {O.}~\bibnamefont {Farago}},\ }\href@noop {} {\bibfield  {journal} {\bibinfo
  {journal} {J. Chem. Phys.}\ }\textbf {\bibinfo {volume} {141}},\ \bibinfo
  {pages} {194108} (\bibinfo {year} {2014})}\BibitemShut {NoStop}%
\end{thebibliography}%

\end{document}